\documentclass[lettersize,journal]{IEEEtran}
\usepackage{amsmath,amsfonts}
\usepackage{algorithmic}
\usepackage{algorithm}
\usepackage{array}
\usepackage[caption=false,font=normalsize,labelfont=sf,textfont=sf]{subfig}
\usepackage{textcomp}
\usepackage{stfloats}
\usepackage{url}
\usepackage{enumerate}
\usepackage{verbatim}
\usepackage{graphicx}
\usepackage{xcolor}
\usepackage{cite}
\usepackage{tabularx,booktabs,textcomp}
\usepackage{amsthm}

\usepackage{amsmath,amssymb,amsthm}

\usepackage{amsmath,amssymb,amsthm}

\makeatletter 
\newcommand{\linebreakand}{%
  \end{@IEEEauthorhalign}
  \hfill\mbox{}\par
  \mbox{}\hfill\begin{@IEEEauthorhalign}
}
\makeatother 

\begin{document}



\title{Advancements in Arc Fault Detection for Electrical Distribution Systems: A Comprehensive Review from Artificial Intelligence Perspective}

\author{{Kriti Thakur, Divyanshi Dwivedi, K. Victor Sam Moses Babu, Alivelu Manga Parimi, Pradeep Kumar Yemula, Pratyush Chakraborty, Mayukha Pal}
 

\thanks{(Corresponding author: Mayukha Pal)}

\thanks{Ms. Kriti Thakur is a Data Science Research Intern at ABB Ability Innovation Center, Hyderabad 500084, India, and also a Ph.D. Research Scholar at the Department of Electrical and Electronics Engineering, BITS Pilani Hyderabad Campus, Hyderabad 500078, IN.}
\thanks{Mrs. Divyanshi Dwivedi is a Data Science Research Intern at ABB Ability Innovation Center, Hyderabad 500084, India, and also a Research Scholar at the Department of Electrical Engineering, Indian Institute of Technology, Hyderabad 502205, IN.}
\thanks{Mr. K. Victor Sam Moses Babu is a Data Science Research Intern at ABB Ability Innovation Center, Hyderabad 500084, India and also a Ph.D. Research Scholar at the Department of Electrical and Electronics Engineering, BITS Pilani Hyderabad Campus, Hyderabad 500078, IN.}
\thanks{Dr. Alivelu Manga Parimi is a Professor in the Department of Electrical and Electronics Engineering, BITS Pilani Hyderabad Campus, Hyderabad 500078, IN.}
\thanks{Dr. Pradeep Kumar Yemula is an Assoc. Professor with the Department of Electrical Engineering, Indian Institute of Technology, Hyderabad 502205, IN.}
\thanks{Dr. Pratyush Chakraborty is an Asst. Professor in the Department of Electrical and Electronics Engineering, BITS Pilani Hyderabad Campus, Hyderabad 500078, IN.}
\thanks{Dr. Mayukha Pal is with ABB Ability Innovation Center, Hyderabad-500084, IN, working as Global R\&D Leader – Cloud \& Analytics (e-mail: mayukha.pal@in.abb.com).}
}


\maketitle

\begin{abstract}

This comprehensive review paper provides a thorough examination of current advancements and research in the field of arc fault detection for electrical distribution systems. The increasing demand for electricity, coupled with the increasing utilization of renewable energy sources, has necessitated vigilance in safeguarding electrical distribution systems against arc faults. Such faults could lead to catastrophic accidents, including fires, equipment damage, loss of human life, and other critical issues. To mitigate these risks, this review article focuses on the identification and early detection of arc faults, with a particular emphasis on the vital role of artificial intelligence (AI) in the detection and prediction of arc faults. The paper explores a wide range of methodologies for arc fault detection and highlights the superior performance of AI-based methods in accurately identifying arc faults when compared to other approaches. A thorough evaluation of existing methodologies is conducted by categorizing them into distinct groups, which provides a structured framework for understanding the current state of arc fault detection techniques. This categorization serves as a foundation for identifying the existing constraints and future research avenues in the domain of arc fault detection for electrical distribution systems. This review paper provides the state of the art in arc fault detection, aiming to enhance safety and reliability in electrical distribution systems and guide future research efforts.

\end{abstract}

\begin{IEEEkeywords}
Arc fault detection, Electrical distribution, Artificial intelligence, Machine learning, Neural networks.
\end{IEEEkeywords}

\section{Introduction}
\label{section:Introduction}

Electrical energy plays a crucial role in the world's economy, serving as a key factor in the quality of living within a society.  It empowers the movement of both goods and people, fuels manufacturing operations, facilitates diverse communication methods, and ensures uninterrupted data storage – all of which are vital components of the modern economy. Additionally, it plays an indispensable role in advancing medical diagnostics, optimizing hospital management, and supporting critical functions in airports and other essential sectors, ultimately enhancing human capabilities. In the current era, there is a growing worldwide consumer demand for power utilities to provide power supply characterized by improved reliability and superior quality. The complexity of the electrical distribution system is steadily increasing due to the greater integration of distributed energy resources \cite{Dwivedi2022AMF,iet:/content/conferences/10.1049/icp.2023.0282}.
 Thus, electricity has transitioned from being a premium commodity to a more accessible and commonplace resource. It has become an important resource with significant economic and security significance. However, power outages and blackouts represent significant disruptions in the continuous functioning of essential services, posing critical interruptions to daily life and routine activities.

Fires pose a significant risk to the safety and well-being of human beings. Electrical fires are a commonly encountered category of fires, comprising an average of 30.2\% of all fire incidents and surpassing 50\% of all incidents classified as big fires \cite{si2014analysis}. By statistical data obtained from fire administrations, it has been observed that electrical fires commonly arise due to several factors such as arc fault, over current, short circuit, and leakage \cite{si2014analysis,vadoli2011lessons}.
 According to the findings of the United States Fire Administration (USFA), the sources of heat in electrical fires occurring in residential buildings, it was determined that arc fault was responsible for 82\% of these incidents \cite{propertiesunited}. Thus, arc faults are widely recognized as a significant factor contributing to electrical fires \cite{propertiesunited,lee2000new}. According to data provided by the USFA, an approximate total of 372,900 incidents of residential fires in buildings occurred annually in the United States between 2011 and 2013. These fires resulted in approximately 13,125 injuries, 2530 accidental deaths, and property damage amounting to 7 billion dollars \cite{propertiesunited}. In the United States, an annual average of about 1700 individuals suffer fatalities or sustain injuries as a result of electrical fires \cite{usf}. In the year 2019, the Korean government released a report discussing the factors contributing to 23 fire incidents in energy storage systems (ESSs). The insufficiency of electrical protection mechanisms in these systems, particularly in terms of safeguarding against Direct Current (DC) arc faults, is highlighted \cite{han2022spectrum}. 
  The key contributions of the work are as follows:

\begin{itemize}
    \item A thorough review and analysis of the latest advancements in the field of arc fault detection for electrical distribution systems. This includes a comprehensive overview of emerging technologies and methodologies.
    \item The importance of early detection and identification of arc faults to prevent potential catastrophic accidents, such as fires, equipment damage, and loss of life. It emphasizes the proactive role of arc fault detection in mitigating risks.
    \item The role of AI in the detection of arc faults. It provides the superior performance of AI-based methods in accurately identifying arc faults, providing a detailed examination of how AI contributes to enhanced detection capabilities.
    \item Identification of existing gaps in arc fault detection technologies and provides a basis for future research directions.
\end{itemize}

The review is organized as follows: Section \ref{section:Introduction} provides the background of arc faults and their effect on the surroundings and why we have to detect arc faults. Section \ref{sec:arc_overview} provides an overview of an arc fault. Section \ref{section:outline_arc_methods}  presents the overview of arc fault detection methods. Section \ref{section:challenges_opportunities} provides the challenges of the traditional methods and opportunities of AI-based methods and the comparison from the literature. Finally, in Section \ref{section:Conclusion}, conclusions on the comprehensive exploration are provided with a discussion on the future scope for arc fault detection.

\section{Arc Fault Overview}
\label{sec:arc_overview}

As per the guidelines outlined in the UL Standard UL1699 \cite{UL1699}, an arc fault is characterized as the sustained release of electrical energy in the form of light through an insulating medium, which is frequently followed by the partial vaporization of the electrodes. The occurrence of this phenomenon could be attributed to several factors, including the deterioration of the insulating medium inside the electrical circuit, the weakening of electrical connections, the presence of humidity in the surrounding air, and a decrease in the insulation strength of the conducting cable. The elevated temperature resulting from the occurrence of an arc fault discharge phenomenon possesses a strong tendency to initiate combustion in the adjacent combustible materials, hence constituting a notable contributing factor to the incidence of electrical fires. The increasing reliance of societies on energy and the growing interconnectedness of their infrastructures necessitate the prioritization of safe and reliable functioning of distribution grids. This entails the development of fault detection systems to limit the consequences of power outages/electrical fires. Therefore, arc fault detection is of the utmost importance.

\begin{figure}[htb]
    \centering
    \includegraphics[width=6.5cm]{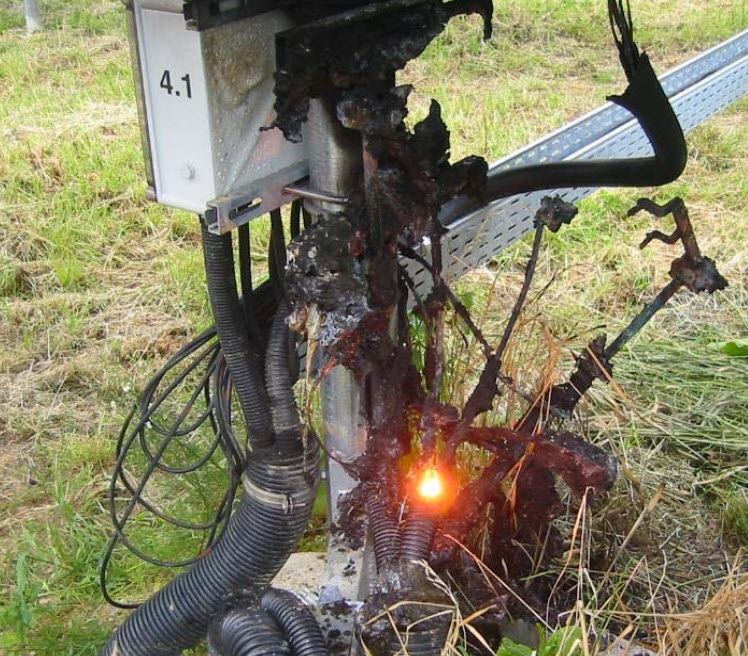}
    \caption{An illustrative image of a practical arc fault persisting within the combiner box \cite{osti_1671730}.}
    \label{fig:FIG5}
\end{figure}

 The occurrence of an electric arc phenomenon arises when a short circuit happens, resulting in the flow of a substantial amount of current. As the contacts of the circuit breaker separate, the contact area diminishes, leading to an increase in current density and a subsequent elevation in temperature. This rise in temperature causes the ionization of the surrounding air.
 The ionized air acts as a conductor between the contacts, and an arc is struck between the contacts. Common forms of damage often observed in such scenarios encompass the occurrence of excessive heat in copper wiring, the liquefaction of aluminum rods, and the release of hazardous gases. The occurrence of overheating could lead to hazardous elevations in pressure, hence resulting in the potential explosion of switchgear. 
 An illustrative image of a practical arc fault persisting within the combiner box is depicted in Figure \ref{fig:FIG5}.
\begin{figure*}[htp]
    \centering
    \includegraphics[width=12cm]{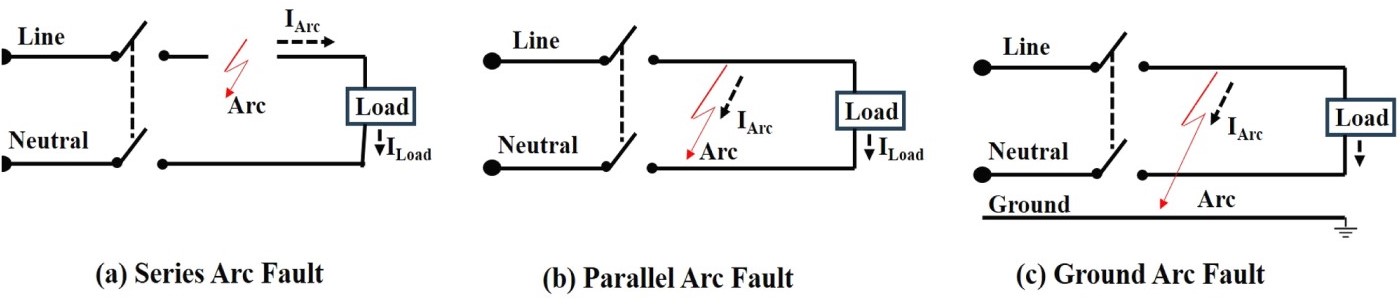}
    \caption{Various types of arc faults.}
    \label{fig:FIG22}
\end{figure*}

 Arc fault could be classified into three types: series arc fault, parallel arc fault, and ground arc fault, depicted in Figure \ref{fig:FIG22} \cite{artale2017arc}. A series arc fault pertains to conductor disconnection in power transmission lines, while a parallel arc fault occurs due to insulation breakdown between two or more parallel lines caused by external forces or heat. A ground arc fault occurs when an unintentional electrical connection is established between an electrical conductor, such as a wire, and the earth's surface. This situation may result in an abnormal current flow, which has the ability to cause harm to equipment, spark fires, or pose dangers to electrical systems. Distinguishing between an arc fault and a normal signal is vital in the context of arc fault detection. It is imperative to find the presence of elevated current and voltage levels, as well as high-frequency fluctuations, together with the accompanying manifestations of sound, vibration, light, and electromagnetic energy that are indicative of fault arcs \cite{lu2020dc,xiong2017novel}. The energy generated during an arc fault, according to Joule's law, is:
 
\begin{equation}
    \Delta E = R \, I^2 \Delta t
\end{equation}

where \( R \) is the overall current path's resistance which includes the arc impedance, \( I \) represents the fault current, and \( \Delta t \) represents the period when the fault exists.
 
\begin{figure*}[htp]
    \centering
    \includegraphics[width=10cm]{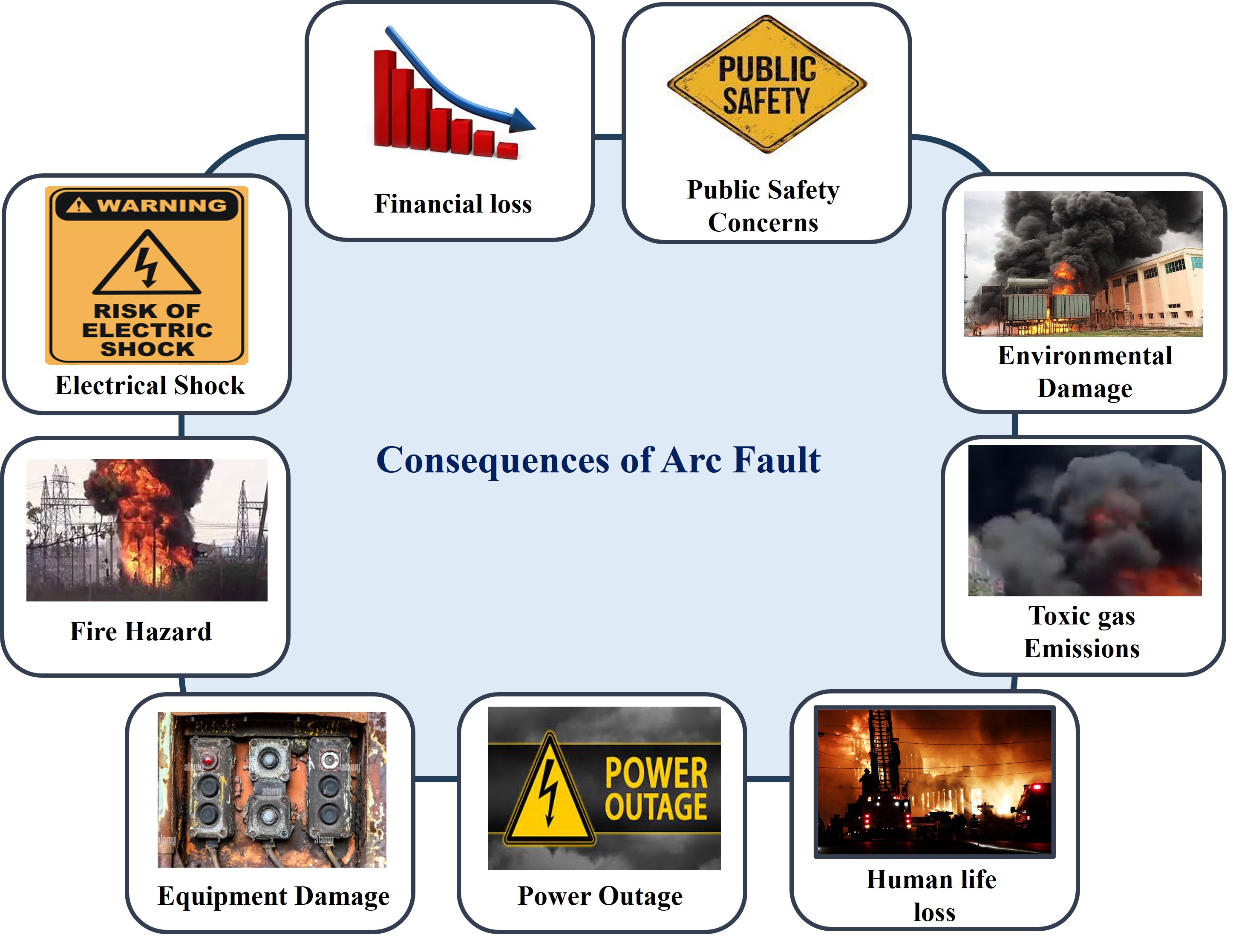}
    \caption{The most significant risks associated with arc faults.}
    \label{fig:FIG2}
\end{figure*}

 Notably, series arc faults are the most commonly occurring type \cite{lin2004new,naidu2006arc,gregory2004more}. Series arc faults are considered as an additional resistance inside the electrical system, resulting in a reduction of the arc fault current. As a result, traditional protective mechanisms are unable to be activated. If series arc faults are not promptly discovered and mitigated, they have the potential to impact the associated circuits within the system, cause damage to the electricity supply channels and system controllers, and potentially result in blasts \cite{chen2019time,guo2018series}. Arc faults in electrical systems may cause substantial risks, giving rise to a range of safety considerations and severe harm. A comprehensive awareness of these potential dangers is important to effectively build, uphold, and manage electrical systems with utmost safety. The most significant risks associated with arc faults are illustrated in Figure \ref{fig:FIG2}.

\subsection{Arc Fault Characteristics}

\begin{figure}[htp]
    \centering
    \includegraphics[width=10cm]{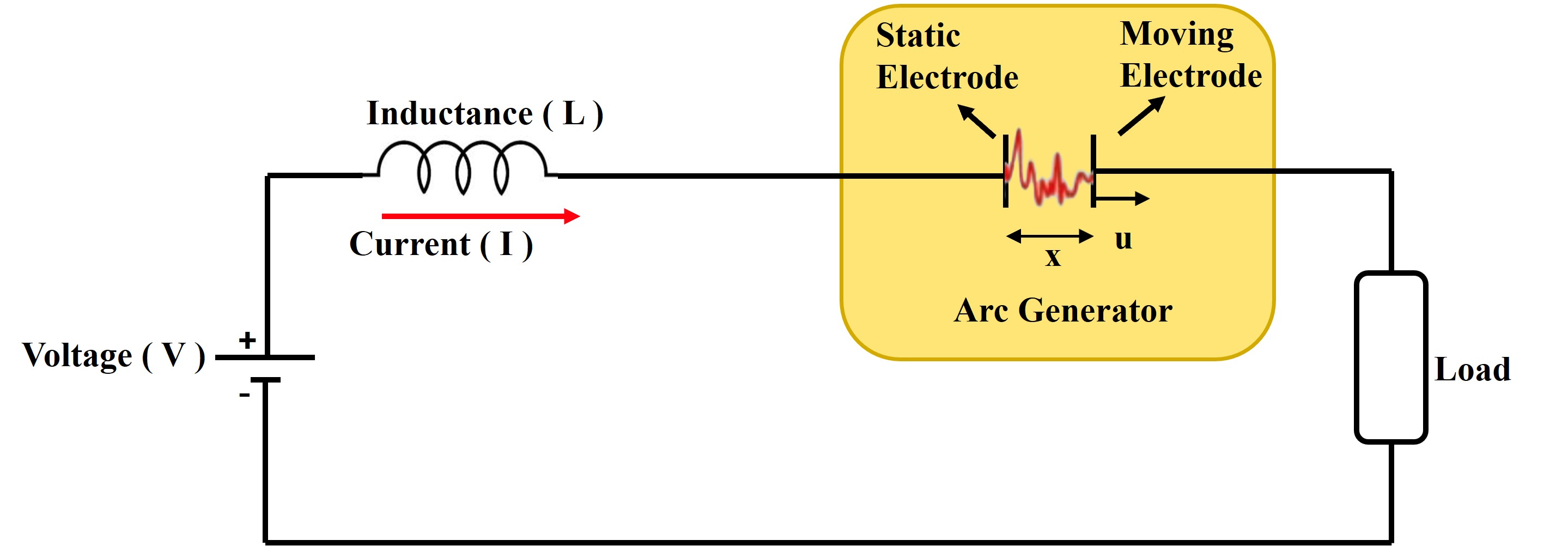}
    \caption{ The equivalent electrical model representing a DC series arc fault.}
    \label{fig:FIG3}
\end{figure}

The electrical behavior of arc faults is inherently complicated, chaotic, and unpredictable, and it is easily
understood and represented using an equivalent circuit model \cite{gattozzi2012analytical,xiong2018electromagnetic,ammerman2009dc}. The circuit depicted in Figure \ref{fig:FIG3} exhibits a gap inside the conduction route, which is connected in series with a line inductance denoted as $L$. Additionally, the circuit has a load that is powered by a  (DC) voltage source represented by $V$. The distance $x$ denotes the gap separation between an electrode that remains fixed and the other electrodes that move away at a velocity $u$, which varies with time. Upon the first separation at the gap, it is believed that an arc is formed, characterized by an arc voltage \(v_{arc}\)  and an arc current \(i_{f}\).
The primary parameters utilized to depict the electrical properties of an arc and assess their influence on the power system in which they manifest are the arc voltage, current, and resistance.

\begin{itemize}
    \item \textbf{Arc Voltage:} Arc voltage refers to the electrical potential difference that develops between the boundaries of the gap. The initiation of an arc necessitates a minimum voltage, which subsequently escalates until the arc is terminated. The voltage stability and sustainability of an arc are dependent upon the length of the gap. If the gap length is within an appropriate range, the voltage will stay in stable condition, hence facilitating a prolonged arc. Conversely, if the gap length exceeds a certain threshold, the arc will undergo self-extinguishment.

\end{itemize}
\begin{itemize}
    \item \textbf{Arc Current:} The arc current refers to the electric current that passes through the gap. The behavior of arc current is inversely related to arc voltage. Upon initiation of the arc, the current experiences an abrupt fall. As the arc voltage increases, the current continues to decline until the arc is sustained. Conversely, if the arc has been quenched, the current will reach zero.

\end{itemize}
\begin{itemize}
    \item \textbf{Arc Resistance:} The behavior of a series arc exhibits resistive characteristics, characterized by a highly nonlinear arc resistance due to the unpredictable nature of the arc phenomena. The arc resistance could be determined by measuring the arc voltage and current at various time intervals. This calculation is commonly performed by averaging the arc resistance across the stable burning duration.
\end{itemize}

\subsection{Arcing Fault model}

\begin{figure}[h!]
    \centering
   \includegraphics[width=8.75cm]{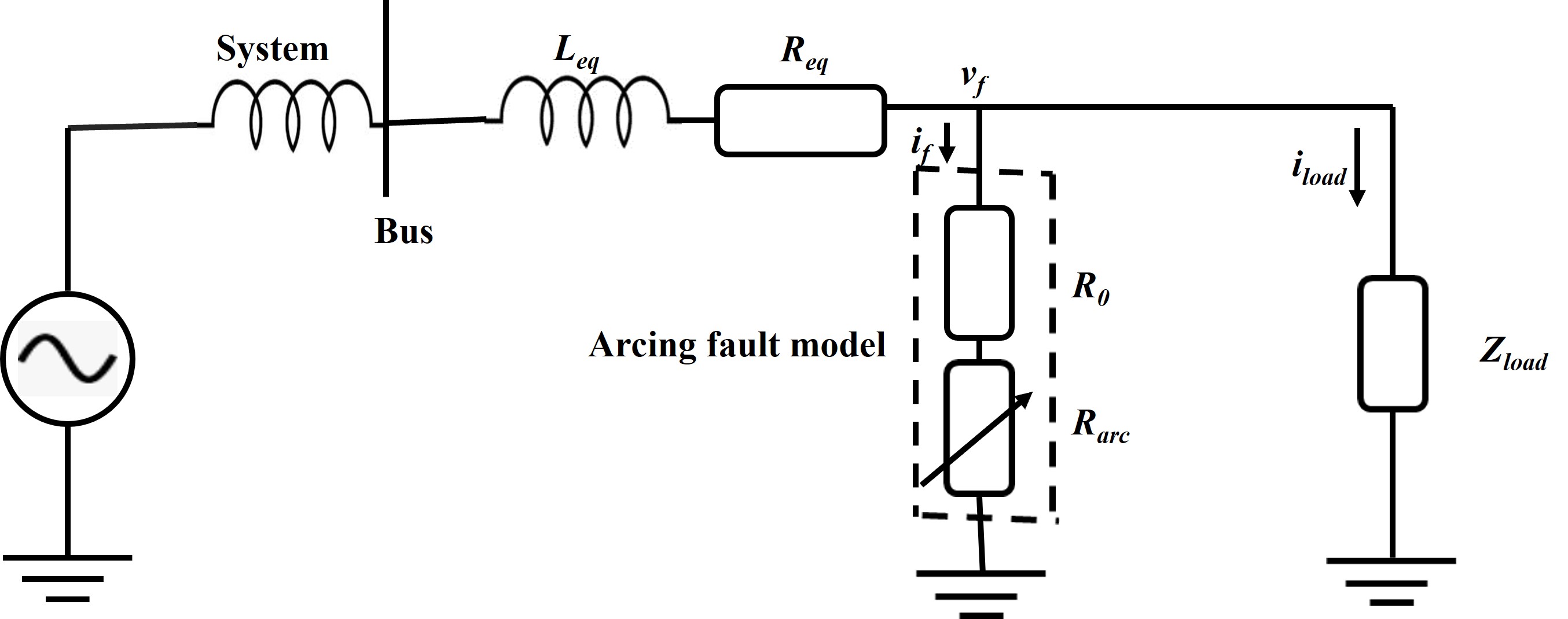}
    \caption{The distribution system's general arcing fault structure \cite{zhang2016model}.} 
    \label{fig:FIG4}
\end{figure}

In the context of a distribution system, it is possible to represent a branch that experiences an arcing fault by employing a model wherein an arc is connected in series with a constant resistance $R_0$ \cite{elkalashy2007modeling}. The general arcing fault model in the distribution system is shown in Figure \ref{fig:FIG4}, where \(R_{arc}(t)\) is the time-varying arc resistance. Let \(g_{arc}(t) = \frac{1}{R_{arc}(t)}\) be the arc conductance, which is also time-varying. Thus, if \(i_f(t)\) is the arc current, the voltage across the arc could be expressed as:

\begin{equation}
    v_f(t) = i_f(t) \cdot R_0 + \dfrac{i_f(t)}{g_{arc}(t)} \label{eq:vf}
\end{equation}

The utilization of an arc model (AM) is necessary in order to accurately represent the function \(g_{arc}(t)\) for the purpose of modeling the detection challenge. Extensive study has been carried out on AMs over the course of a couple of decades, resulting in the proposal of numerous AMs to address various motives, such as the analysis of circuit-breaker arcs \cite{darwish2005universal}, arc energy calculation \cite{parise2013simplified}, and arcing fault analysis \cite{torres2014modeling,ospina2008analysis}.

The AM that connects \(g_{arc}(t)\)  and the current of the arc branch \(i_f(t)\) is as follows:

\begin{equation}
    \dfrac{d g_{arc}(t)}{dt} = \dfrac{1}{\tau} \left( \dfrac{ |i_f(t)|}{u_0 + r_0 \, |i_f(t)|} - g_{arc}(t) \right) \label{eq:garc}
\end{equation}

where \(u_0\) is the characteristic arc voltage, \(r_0\) is the characteristic arc resistance, and \(\tau\) is the arc time constant.

Various types of loads have a significant impact on arc fault currents. The fault currents generated by specific household appliances exhibit minimal variations, and the characteristics of arc faults are typically obscured by load currents and ambient noise. Finding a fault feature that consistently performs for all types of loads poses a significant challenge. In addition, it should be noted that the home the electrical norm for single-phase alternating current (AC) in the United States is 120 V (60 Hz), although in other countries such as China, Germany, Switzerland, and Korea, it ranges from 220 V to 240 V (50 Hz to 60 Hz). Arc faults are more likely to result in electrical fires when higher voltages are present\cite{kolker2007study,shea2008comparing,carvou2009electrical,ko2013analysis}. Given a case study in \cite{kolker2007study}, the probability of fire initiation resulting from arc faults exhibits an escalation from 3.5\% for a voltage of 120 V to 83\% for a voltage of 240 V. It is worth noting that the nominal current levels for both scenarios remain constant at 15 A. In contrast to the voltage of 120 V, the elevated voltages ranging from 220 to 240 V have a greater propensity to induce gap breakdown and generate a bigger number of arcs \cite{shea2008comparing,carvou2009electrical}. Arc currents often exhibit a continuous pattern in systems operating at higher voltages, however in 120 V systems, they may occasionally display an intermittent behavior \cite{kolker2007study,shea2008comparing}. Consequently, it could be observed that systems operating at higher voltage levels have the ability to create greater arc energy, thereby creating more favorable circumstances for the initiation of electrical fires \cite{shea2008comparing,carvou2009electrical}.

\subsection{How to Prevent Arc Faults?}
The existence of an arc fault signifies a severe breakdown in the power system. An arc flash incident has the potential to inflict considerable harm on individuals engaged in operations, lead to substantial financial losses stemming from equipment damage, and trigger extended periods of system downtime. The extinction of this arc is achieved through the process of cooling and elongation of the arc. During the procedure of interrupting a circuit, the arc that is formed serves the purpose of dissipating the energy that is stored in the inductance of the circuit. Following the successful suppression of the arc, the dielectric strength commences its healing process, while the circuit breaker successfully separates the fault, under the assumption that no re-striking happens. The categorization of circuit breakers is established based on the particular arc quenching medium utilized and the cooling configuration utilized within the breaker. The following discussion provides a concise analysis of the air arc chute circuit breaker, vacuum interrupter, and SF6 circuit breaker as illustrative examples:

\begin{enumerate}[i.]
    \item Vacuum circuit breakers employ a confined container filled with a vacuum as the medium for extinguishing electrical arcs, owing to the high dielectric strength exhibited by the vacuum. Vacuum circuit breakers are extensively utilized in medium voltage AC distribution grids due to their numerous advantages. These advantages include their self-contained nature and compatibility with the environment, their high reliability and low maintenance requirements, their non-flammability, their compact size and short stroke, their low energy consumption for operation, and their extended service life.
    \item Sulfur hexafluoride (SF6) circuit breakers employ pressurized SF6 gas within a confined container as the medium for extinguishing electrical arcs. Sulfur hexafluoride is introduced in a controlled manner during the arc interruption procedure to aid in the suppression of the electric arc. SF6 is frequently employed in high-voltage AC networks \cite{pei2016review}.
    \item Molded-case circuit breakers (MCCBs) are essential components employed in numerous distribution systems to safeguard against network disturbances \cite{switchgear2006controlgear}. During a fault event, a substantial surge in electric current flows through the terminals of the MCCB, causing the contacts to separate due to the activation of an electromagnetic mechanism. Upon initiation, a low-voltage electric discharge is established between the contact points, resulting in the formation of a high-temperature plasma with temperatures reaching several thousand degrees Celsius \cite{yang2013low}. The presence of plasma induces thermal degradation in the insulating materials and erosion of the contacts, while also emitting high UV radiation that could exacerbate the breakdown of the insulating materials. The production of electrically conductive soot in a polyester-based MCCB has a negative impact on the device's long-term functioning. Arc chutes are employed within the MCCB to elongate or divide the arc and facilitate its cooling through thermal exchange. This aids in the expeditious extinguishing of the arc, hence decreasing the thermal deterioration of the insulating materials and lowering erosion of the contacts. The Arc Vault protection system \cite{4} represents a significant advancement in arc fault prevention technology. It is capable of rapidly interrupting an arcing fault within a timeframe of less than eight milliseconds. This system effectively enhances the mitigation of arc flash hazards, regardless of whether equipment doors are open or closed. The implementation of this rapid arcing fault containment technique has the potential to mitigate the extent of work-related injuries and minimize the loss of productivity .
\end{enumerate}

In contrast to the conventional circuit breaker, the Asea Brown Boveri (ABB)'s solid-state circuit breaker employs power electronic components and sophisticated software algorithms to replace the mechanical components found in classic electromechanical circuit breakers. This substitution enables the solid-state circuit breaker to swiftly regulate power and interrupt current flow. The ABB circuit breaker concept originated by ABB's Research and Development unit situated in Bergamo, Italy. The circuit breaker employs the Insulated Gate Commutated Thyristor (IGCT) semiconductor technology plan, which incorporates an innovative incorporated predictive power management software, protection algorithm, and advanced-level communication \cite{huo2022review}.
\begin{figure}[htp!]
    \centering
    \includegraphics[width=9cm]{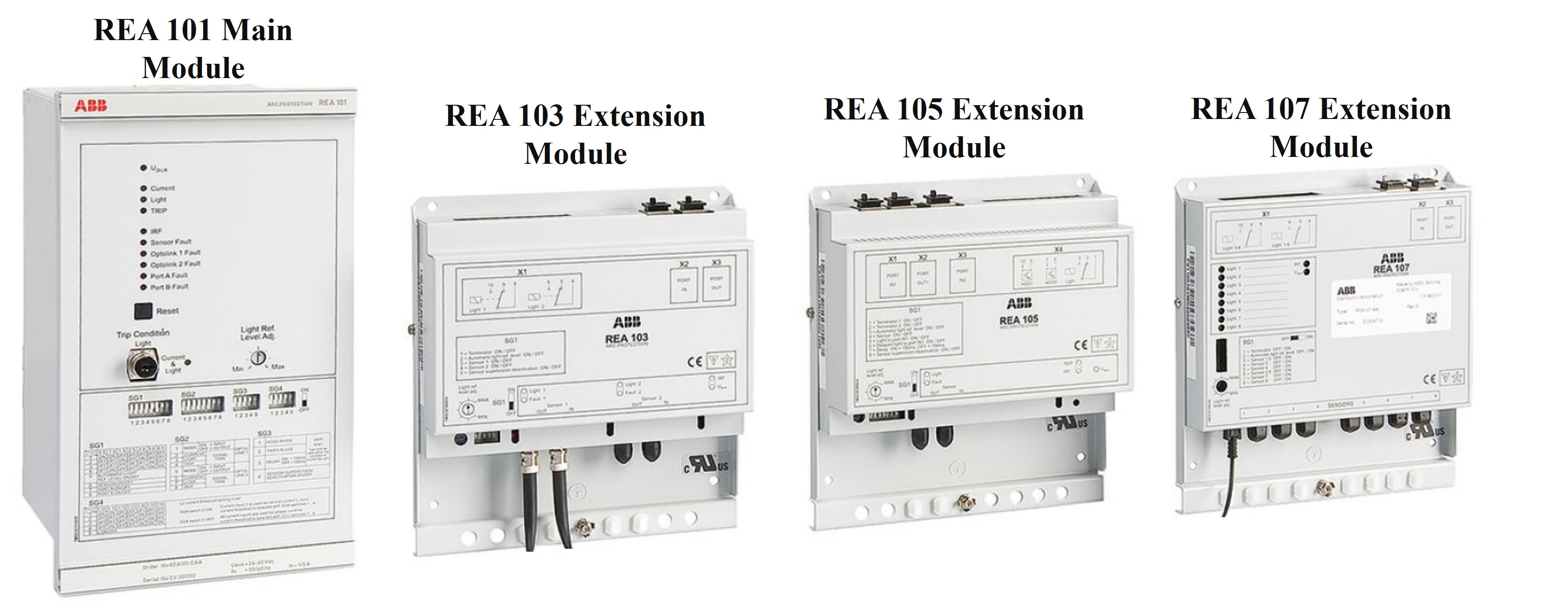}
    \caption{Product image of Arc Fault Protection System \cite{5}.}
    \label{fig:rea}
\end{figure}

The REA system, a product by ABB \cite{5}, is an efficient and adaptable arc fault protection system designed for air-insulated switchgear operating at low voltage and medium voltage levels; the image is illustrated in Figure \ref{fig:rea}. The REA 101 main module functions autonomously or in conjunction with other REA 101 modules and extension modules. The REA 103 extension module expands the arc flash detection area by incorporating two sensor fibers arranged either in a loop or radially. Meanwhile, the REA 105 extension module identifies arcs using a loop-type or radial sensor fiber configuration and offers two high-speed semiconductor outputs for circuit breaker tripping. Additionally, it serves as an overcurrent information link between two REA 101 main modules. Lastly, the REA 107 extension module is designed for arc detection, employing eight lens sensors arranged radially.

The integration of a rapid and discerning arc fault protection system is an inherent component of contemporary switchgear panels. Furthermore, it serves as a prudent investment in safety and security for antiquated switchgear, as it effectively safeguards human lives and mitigates the risk of asset impairment.

The REA arc fault protection system employs two distinct types of sensors to detect light. The first type is a non-shielded, bare-fiber sensor that is capable of sensing light across its whole length. The second type consists of light-collecting lens-type sensors, with normally one sensor installed in each switchgear compartment \cite{5}. 

The functionality of the REA arc fault protection system is predicated on the identification of either the intense luminosity emitted by an electric arc in isolation or the detection of luminosity in conjunction with concurrent phase or neutral overcurrent. Upon the identification of an arc fault, the REA arc fault protection mechanism promptly issues trip commands to all circuit breakers responsible for supplying power to the area affected by the fault, accomplishing this task within a time frame of less than 2.5 milliseconds. In addition, the operational indications of the REA arc fault protection system aid in the localization of problems by providing selective guidance to maintenance personnel toward the specific fault zone that has been recognized by the arc fault protection system.

ABB provides active arc fault protection solutions that encompass both arc fault detection and removal, hence mitigating the aforementioned dangers. Collectively, these systems are engineered to identify an internal arc within a time frame of 1.5 milliseconds and subsequently mitigate it within a duration of less than 4 milliseconds. As a result, the enhancement of the electrical system's reliability and security may be greatly increased.

Arc Fault Circuit Interrupters (AFCIs) is an instrument that interrupts a circuit if it recognizes a dangerous electrical arc by recognizing the patterns of electrical arcs. Individual circuits are protected by these, which are generally installed in household and commercial electrical panels. AFCIs help in protecting both parallel and series arcs in connected loads and the wiring \cite{martel2010study}. AFCIs rely on current as a measurement parameter due to its ease of measurement and the absence of a need to modify the circuit or wire connections. Time-domain characteristics are retrieved to detect arcs.
Arc faults refer to unintentional electrical arcs that have the potential to ignite combustible objects within a residential structure. The AFCI is characterized by its compact size, ease of installation, and its ability to effectively handle shared and mixed neutrals. The AFCI complies with the National Electrical Code of 2020 \cite{1}.

\subsection{Why Fault Detection is Important?}
The timely identification of faults, whether in equipment or the intermediate products of manufacturing processes, carries substantial importance. This is vital to uphold the standards of the end product, prevent the occurrence of abnormal operating circumstances, mitigate costly repairs, and avert potential shutdowns in the production process. Utilizing a typical method for network operation, characterized by an operating period ranging from 80 to 100 milliseconds, leads to the occurrence of cable fire as well as the melting of copper and steel components \cite{6}. The increasing intricacy of industrial systems and the expanding volume of accessible data have stimulated the advancement of intelligent systems for automated fault prediction and detection. These systems mostly rely on industry 4.0 technologies, with a special emphasis on deep learning techniques. The presence of a substantial and uninterrupted stream of data, along with the diverse nature and fluctuating characteristics of the data at hand, poses significant challenges in employing traditional data analysis methods for decision-making purposes, such as in the domain of fault risk assessment. Consequently, there has been an increasing inclination in the development of data analysis systems that utilize AI approaches \cite{zeba2021technology}.

\begin{figure*}[htp]
    \centering
    \includegraphics[width=12cm]{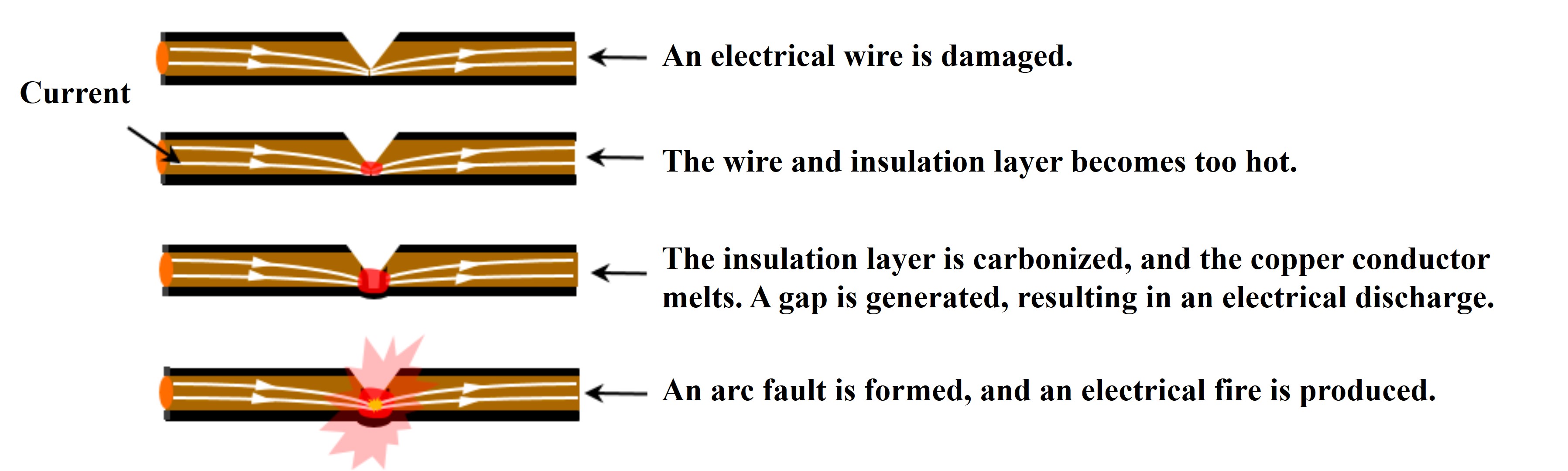}
    \caption{An electrical fire that arises as a consequence of an arc fault.}
    \label{fig:FIG6}
\end{figure*}

The application of machine learning and supervised learning algorithms in fault recognition, diagnosis, and failure prediction has experienced significant advancements over the past few decades due to the adoption of digital technologies and the availability of data. The aforementioned technological improvements have facilitated the development of precise models that support the formulation of effective and sustainable policies aligned with the strategic goals of industries. Consequently, these advancements have improved the decision-making process. Hence, it is imperative to precisely identify series arc problems to maintain electrical safety. Nevertheless, the detection of series arc faults is a major obstacle in actual engineering applications due to their inherent properties, including high level of unpredictability \cite{humbert2021serial}. Figure \ref{fig:FIG6} depicts the sequential stages involved in the occurrence of an electrical fire resulting from an arc fault. 

The insulating layer of the compromised electrical wire undergoes carbonization, leading to the following occurrence of random arcs. With time, the carbonized pathways undergo further expansion, resulting in the formation of visible arcs. The aforementioned arcs have the capacity to discharge an immense quantity of energy and could attain temperatures as high as 13,000 K \cite{2}. The presence of combustible objects in close proximity facilitates the generation of a fire with relative ease. The primary objective of fault diagnosis is to efficiently identify the precise locations of faults. Once faults are detected, they must be appropriately classified to facilitate fault localization. Technical processes need advanced supervision and fault identification to enhance safety measures and reliability \cite{isermann1997supervision}.

\begin{figure}[htp]
    \centering
 \includegraphics[width=9cm]{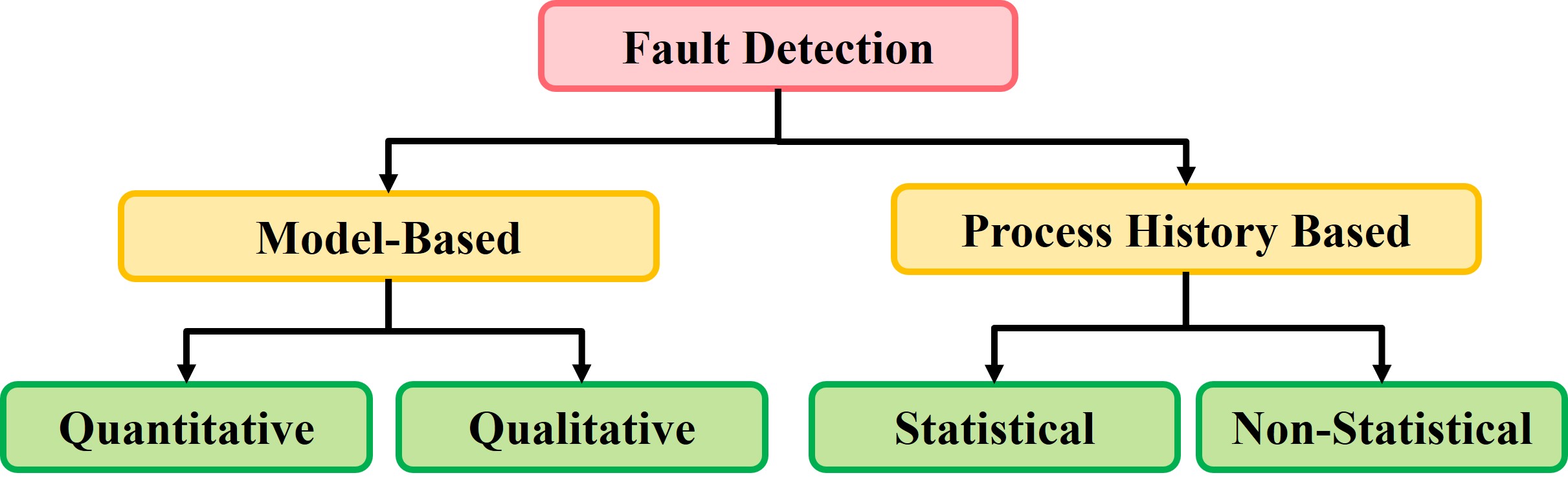}
    \caption{Various fault detection techniques.}
    \label{fig:FIG7}
\end{figure}

Understanding the techniques employed for fault detection is of paramount significance as it enables the timely identification of faults, hence preventing potential harm to the system. The methodology for fault diagnosis is presented in Figure \ref{fig:FIG7}. There are two distinct approaches for fault detection: the model-based method, which is the traditional approach, and the process history-based method, which is recognized as an AI-based approach \cite{katipamula2005methods, article}. 

In order to employ model-based approaches, it is necessary to have an appropriate mathematical model that accurately represents the process system. The aforementioned definition, or pre-existing knowledge, is primarily rooted in the fundamental principles of physics and can encompass both quantitative and qualitative aspects. The traditional model-based method is carried out in two different ways:
\begin{enumerate}
    \item Quantitative Methods: These methods use mathematical modeling of physical processes. These methods, assisted by computational advances, were developed to handle increasingly complex systems and use analytical redundancy by deriving precise mathematical descriptions.
    \item Qualitative Methods: These fault diagnosis methodologies encompass model-based or knowledge-based approaches that rely on heuristic symptoms rather than analytical descriptions to ascertain the present condition of the system. 
\end{enumerate}
In the context of process history-based techniques, the approach relies on a substantial collection of experimental input/output data and endeavors to establish a functional relationship solely based on these characteristics.

\begin{enumerate}
    \item Statistical Methods: Linear regression, multiple regression, polynomial regression, principal component analysis, partial least squares, and logistic regression.
    \item Non-Statistical Methods: Artificial neural networks (ANNs) and various other pattern recognition techniques.
\end{enumerate}

In the realm of arc fault detection, various mathematical models of arcs have been documented in the literature, and this discussion will outline some of their fundamental aspects.
Mathematical models are used to depict the dynamic variations in the series arc using mathematical equations. These models encompass the Cassie, Mayr, Habedank, and several other enhanced models \cite{9605243,habedank1993application} and are known as physics-based AMs.
The Cassie and Mayr AMs are highly renowned and extensively employed AMs that were established based on the notion of energy conversion. The Cassie AM could be mathematically represented as follows:

\begin{equation}
    \dfrac{1}{ g} \dfrac{dg}{dt} = \dfrac{1}{\tau } \bigg(\dfrac{u_{arc} \, i_{arc}}{U^{2}_c} - 1\bigg) \\
\end{equation}

\begin{equation}
    g = \dfrac{i_{arc}}{u_{arc}}
\end{equation}

The variables in the equation are as follows: $g$ denoted as represents arc conductance. $u_{arc}$ denoted as represents arc voltage. $i_{arc}$ denoted as represents arc current. $U_c$ denoted as represents constant arc voltage. $\tau$ denoted as represents the arc time constant. In this particular model, Cassie assumed that the power loss was due to forced convection. This assumption implies a straight proportionality between the area of the arc cross-section and the arc current. Therefore, Cassie's arc model is more suited for simulating events that include substantial amounts of arc current.

In contrast, the Mayr AM is more appropriate for low current arcs due to the assumption made by the Mayr that power loss is primarily attributed to thermal conduction and remains constant \cite{gustavsson2004evaluation}. The equation is illustrated as follows:

\begin{equation}
    \dfrac{1}{g} \dfrac{dg}{dt} = \dfrac{1}{\tau } \bigg(\dfrac{i_{arc}^2}{P_{loss}}\bigg)
\end{equation}

The variables in the equation are as follows: $P_{loss}$ denoted as power loss The value of the unknown constant in equation (6) could be found by analyzing and interpreting experimental data \cite{ju2016arc}.

AMs of Cassie and Mayr are connected in series and make the Habedank model and it is applicable to both large and small currents; the model includes numerous parameters that are challenging to simulate. The equation of the Habedank model  is illustrated as follows:

\begin{equation}
    \dfrac{1}{ g_c} \dfrac{dg_c}{dt} = \dfrac{1}{\tau_c } \bigg[ \bigg(\dfrac{u_{arc} \, g}{U_c \, g_c}\bigg)^2-  1\bigg]  
\end{equation}

\begin{equation}
    \dfrac{1}{ g_m} \dfrac{dg_m}{dt}  = \dfrac{1}{\tau_m} \bigg(\dfrac{u_{arc}^2 \, g^2}{P_{loss} \, g_m} - 1\bigg)
\end{equation}

\begin{equation}
    \frac{1}{g}= \frac{1}{g_c}+\frac{1}{g_m}       
\end{equation}

The variables in the equation are as follows: $g_c$ denoted as represents the arc conductance of the Cassie model, $g_m$ denoted as represents the arc conductance of the Mayr model, $\tau_c$ denoted as represents the arc time constant of Cassie model, $\tau_m$ denoted as represents the arc time constant of Mayr model.

Due to their arbitrary and aperiodic characteristics, mathematical models are unable to adequately characterize the current waveforms resulting from series arc faults in electrical systems under various conditions, as estimated by these models. Hence, the methodologies grounded in mathematical models predominantly remain within the boundaries of simulation.

The study of process history involves the examination and analysis of past events and developments, with a particular focus on the processes and mechanisms that these models are well-suited for situations in which training data are abundant or where the cost of creating or collecting such data is low. The mapping functions have been comprehensively studied and considerably refined. Extensive study has been undertaken in the aforementioned area of pattern identification algorithms, specifically in the domain of machine learning. The aforementioned approaches exhibit significant resilience to system noise in dynamic scenarios.

\section{An outline of arc fault detection approaches}
\label{section:outline_arc_methods} 
\begin{figure*}[htp]
    \centering
    \includegraphics[width=12cm]{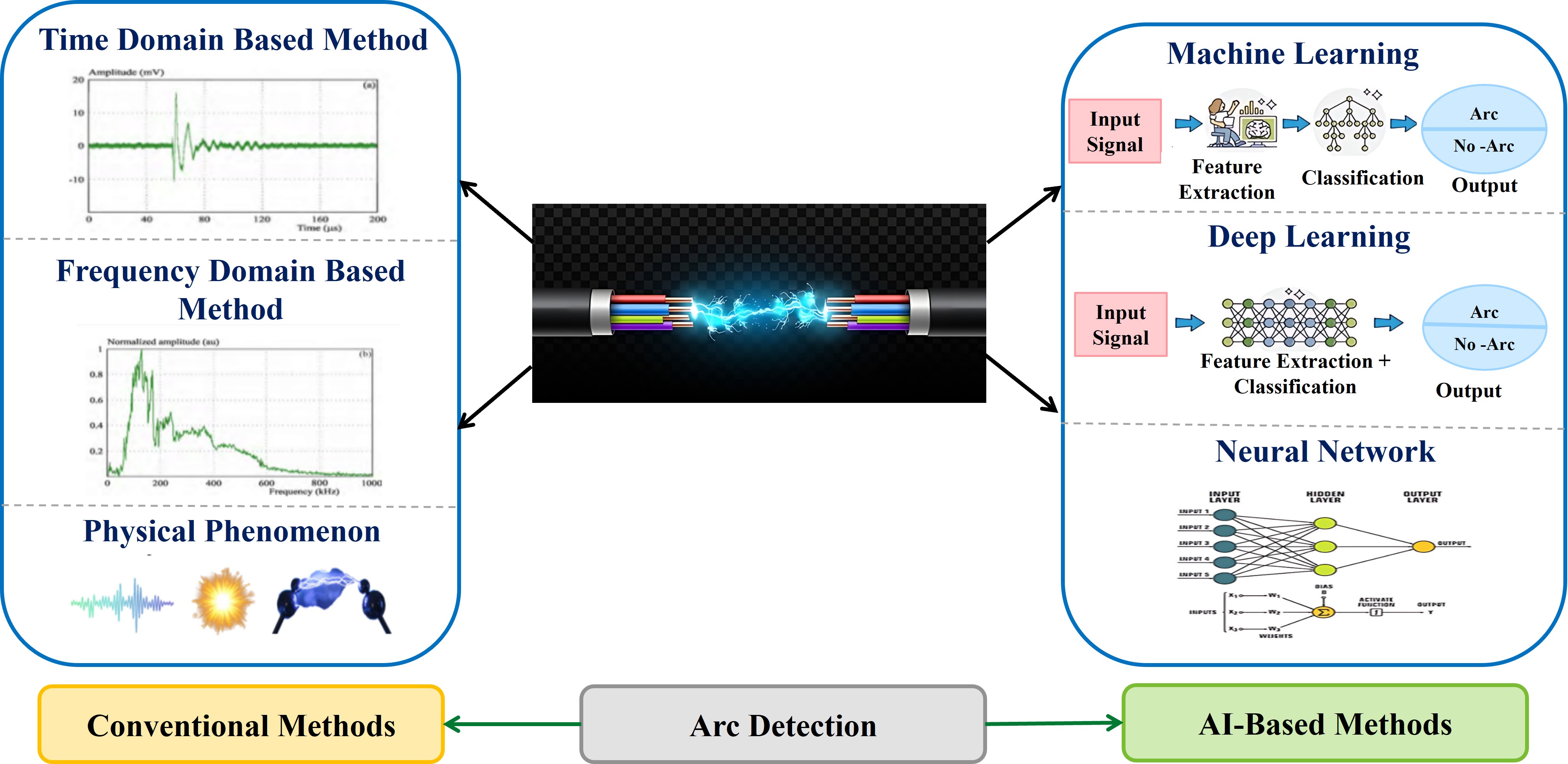}
    \caption{Outline of arc fault detection methods.}
    \label{fig:FIG9}
\end{figure*}

The present study classifies arc fault detection systems into two primary categories: conventional approaches and methods based on AI. The predominant approaches for detecting arc faults involve the study of the time domain, frequency domain, time-frequency domain, or the observation of physical phenomena such as sound, light, or electromagnetic radiation. The methodologies employed in this study involve the examination of electrical signals generated by the occurrence of an arc fault, intending to extract and utilize the pertinent information embedded within these signals. The aforementioned data could be employed to determine the presence of an arc fault. In contrast, AI-based approaches employ machine learning algorithms to analyze power system data to detect arc faults. The aforementioned technologies have demonstrated notable levels of accuracy and efficiency in the identification of arc defects, presenting a potentially advantageous substitute for traditional approaches. Figure \ref{fig:FIG9} illustrates an outline of various techniques employed for arc fault detection.

\subsection{Traditional Arc Fault Detection Approaches}
Multiple approaches use different ideas to identify arc problems, including analyzing electrical current and voltage waveforms or examining arc physical characteristics. Arc fault detection techniques that utilize an electrical quantity could be categorized into two different categories: techniques that rely on current characteristics and techniques that rely on voltage characteristics \cite{liu2017research}. The foundations of conventional arc fault detection techniques are the electromagnetic and thermal fingerprints of electrical arcs.  These techniques are frequently employed to locate hazardous arc faults in electrical systems, which helps to avoid fires and property damage. Typical conventional techniques for arc fault detection include:

\subsubsection{Time Domain-Based Arc Fault Identification Methods} Several detection algorithms have been developed that utilize time-domain properties of current signals, such as shoulder length, change rate, and signal energy \cite{jiang2021coupling,ahmadi2019new}. The chosen time-domain properties, however, are only appropriate for a few particular kinds of loads. In \cite{bao2019novel}, authors introduced the utilization of time domain analysis as a means to identify the happening of arc faults. This technique involves the examination of current and voltage data. { Voltage signals could be significantly affected by drops or fluctuations in the supply voltage, and the characteristics of voltage are also susceptible to variations in circuit types.}
The above-discussed approaches provide rapid detection rates through the utilization of uncomplicated circuits and algorithms. However, they are vulnerable to interference caused by switching noise and fluctuations in load. Several time-domain techniques are commonly employed to identify series arc faults, such as the finite difference method or kurtosis analysis \cite{9756255}. Voltage signals could be significantly affected by drops or fluctuations in the supply voltage, and the characteristics of voltage are also susceptible to variations in circuit types.

\subsubsection{Frequency Domain-Based Arc Fault Identification Methods} In \cite{8395285}, authors employed the tunnel magnetoresistance (TMR) sensor to quantify the alteration in the magnetic field caused by the arc. They further utilized this data to compute the power spectrum density, hence facilitating the detection of the arc. The frequency domain identification of arc fault conditions also used the wavelet packet transform (DPT) and discrete wavelet transform (DWT). The work in \cite{bao2018research} introduces an approach for the detection of arc fault that combines signal collection and higher-order cumulant identification. This approach leverages the electromagnetic coupling phenomenon of high-frequency current. It involves analyzing the coupling signal of the high-frequency current and applying the higher-order cumulant technique when arcing occurs to determine the kurtosis. The study introduces a unified kurtosis threshold value applicable to various scenarios. The proposed approach involves decomposing the frequency response to enhance resolution within the specified frequency range. However, it is crucial to note that this method may amplify the noise signal resulting from the occurrence of an arc fault condition \cite{wang2015arc}. In the study conducted by \cite{syafi2018real}, a detection algorithm was presented that utilizes the cumulative sum of spectrum parity values within the frequency range of 1kHz to 15kHz.
The Fourier transform is a well-established technique utilized for the investigation of frequency domains and is extensively employed in the diagnosis of DC arc faults \cite{johnson2011photovoltaic,johnson2011creating}. The methods described in \cite{calvo2014arcing,lezama2015embedded,lezama2014arc}, utilize correlation coefficients to identify variations in signal form caused by the existence of arc faults. The authors in  \cite{tisserand2015series} utilize algebraic derivatives of line currents as a feature in their study on arc fault detection.
The utilization of wavelet packet technology is employed in \cite{liu2015research} to perform the decomposition, reconstruction, and normalization of the current signal. The investigation of the spectral attributes of arc faults involves the integration of information entropy and short-time Fourier transformation.

In a study conducted by \cite{dolkegowski2021novel}, a method was introduced for the progressive breakdown of a signal over a period of time. This algorithm was found to be approximately seven times more efficient than the Fast Fourier Transform (FFT) and also more cost-effective in terms of implementation for arc fault detection devices. The advent of 5th-generation technology for communication and artificial intelligence has led to significant advancements.

The frequency-domain transformation techniques encompass the discrete Fourier transformation and the chirp zeta transformation. The discrete Fourier transformation is commonly employed because of its capacity to efficiently capture the characteristics of the frequency domain. Nevertheless, the performance of the technique may be influenced by the selected sample frequency and observation window. The wavelet transform is extensively employed for feature extraction and fault detection due to its capacity to offer enhanced local properties of the signal.

\subsubsection{Time-Frequency Domain-Based Arc Fault Identification Methods}
The investigation of novel techniques in the time-frequency domain for the detection of arc faults in electrical systems not only improves accuracy and effectiveness but also broadens its potential applications in various domains. One such domain is biomedical signal processing \cite{Pal2021.09.03.21263098}, where the analysis of dynamic signals is highly valuable, particularly in fields like electroencephalography and electromyography.

The author in \cite{yin2019novel,Pal2021.09.03.21263098}  advocated the use of discrete wavelet transformation for conducting hybrid analysis in the time-and-frequency domain. This study in \cite{en16031256} presents a novel methodology that integrates DWT, empirical mode decomposition (EMD), and dynamic temporal warping (DTW) techniques to address the issue of detecting low-voltage AC series arc faults. The DWT leverages the temporal and spectral features of the arc current data in order to identify the happening of arc faults. The process of EMD involves breaking down the intricate arc fault current signal into a limited number of intrinsic mode signals (IMFs). Subsequently, the instantaneous amplitude of the IMF signal is acquired by the utilization of the Hilbert-Huang transform (HHT). This amplitude serves as a distinguishing characteristic for the identification of arc faults. The outcomes of arc fault detections are contingent upon the outcomes derived from DWT and EMD. In cases where two distinct procedures yield disparate outcomes, the DTW method is employed as a supplementary step to ascertain the presence of an arc fault. This involves utilizing similarity measures between normal and arc fault current waveforms. This study in \cite{wang2018series} presents a novel approach, namely the hybrid time and frequency analysis and fully connected neural network (HTFNN), for the identification of series AC arc faults.
The HTFNN method employs a classification approach to identify the state of samples by categorizing them into three main groups: resistive (Re), capacitive-inductive (CI), and switching (Sw). This categorization is based on the analysis of the fundamental frequency components of the series current. For fine class recognition and state identification, a fully connected neural network (NN) is utilized in each category. This NN is equipped with customized time and frequency indicators as input. The overall accuracies for identifying normal and arcing states in the Re, CI, and Sw categories are 99.64\%, 100\%, and 98.45\%, respectively.

\begin{table}
\centering
\renewcommand{\arraystretch}{1.5}
\caption{Conventional approaches for arc fault detection.}
\begin{tabular}{>{\centering}m{2em} c l}
\toprule
 \multicolumn{1}{>{\centering}m{5em}}{Approach} & 
    \multicolumn{1}{>{\raggedright}m{14em}}{Method} & 
    \multicolumn{1}{>{\centering}m{4em}}{References} \\
\midrule
 \multicolumn{1}{>{\centering}m{5em}} {Time Domain} & 
\multicolumn{1}{>{\raggedright}m{14em}}{Kurtosis Analysis, \\ Finite Difference Method.} & 
\multicolumn{1}{>{\centering}m{4em}}{\cite{jiang2021coupling,ahmadi2019new,bao2019novel,9756255}}\\ 
 \multicolumn{1}{>{\centering}m{5em}} {Frequency domain}  & 
\multicolumn{1}{>{\raggedright}m{14em}}{Discrete Wavelet Transform, \\ Fast Fourier Transform, Discrete Fourier Transformation, and \\ Chirp Zeta Transformation.} & 
\multicolumn{1}{>{\centering}m{4em}}{\cite{8395285,bao2018research,wang2015arc,syafi2018real,dolkegowski2021novel,calvo2014arcing,lezama2015embedded,lezama2014arc,tisserand2015series,liu2015research}}\\
 \multicolumn{1}{>{\centering}m{5em}} {Time-Frequency Domain} & 
\multicolumn{1}{>{\raggedright}m{14em}}{Discrete Wavelet Transform,\\ Empirical Mode Decomposition, \\ Dynamic Temporal Warping, and \\ Hilbert-Huang Transform}  & 
\multicolumn{1}{>{\centering}m{4em}}{\cite{yin2019novel,en16031256,wang2018series,yin2020integrated,wang2014arc,qu2019series,jiang2019series,6558862}}\\
\bottomrule
\end{tabular}\\
\label{tab:traditional_methods}
\end{table}

These approaches enhance the precision of detection; yet, they are unable to attain feature threshold adaption.
The extraction of time-frequency domain aspects of the arc current is commonly performed by researchers through calculation. Subsequently, these features are classified using a classifier\cite{yin2020integrated}. Despite the demonstrated improvements in performance, these methods still need the human extraction of features and their subsequent input into neural networks. The authors of \cite{wang2014arc} conducted a comparative analysis of arc detection approaches utilizing synthetic waveforms, specifically focusing on the performance of FFT and DWT. Their findings suggest that DWT generates detection features that are more readily identifiable.
In \cite{qu2019series,jiang2019series,6558862} use the time-domain and frequency-domain information to find errors.
They are suited for a limited number of load types, and the detection accuracy would decline as the number of load types increased. Table \ref{tab:traditional_methods} presents a comparative analysis derived from previous research articles on conventional approaches utilized for arc fault detection.



 
\subsection{AI-Driven Approaches for Arc Fault Detection: Fundamentals and Methods}

The integration of sophisticated artificial intelligence methodologies, including machine learning, deep learning, visual processing, and recognition of patterns, facilitates the instantaneous examination of extensive datasets produced by electrical systems and the detection of arc faults. In the past couple of years, AI has been extensively utilized in various disciplines, encompassing the field of electrical engineering \cite{chen2020artificial,6547979,battina2015application,qiao2022improving}. The utilization of these AI-based techniques offers clear advantages over conventional arc fault detection methods, particularly in terms of their capacity to process immense quantities of data, high sensitivity in picking small variations in an electrical waveform, and their compatibility with existing electrical systems. Several studies have utilized AI techniques for the purpose of arc fault detection, as illustrated in Table \ref{tab:AI_methods}.

This section presents a brief overview of AI approaches highlighting their significance in the present context and potential topics for future research. The subsequent section of this discussion elucidates the use of AI techniques in arc fault detection, so facilitating the reader's understanding of its enhanced precision and expedited outcomes.

\subsubsection{Machine Learning Based Methods}

\begin{enumerate}[i.]
    \item Support Vector Machines (SVM): The SVM uses Vapnik–Chervonenkis (VC) theory. Next, Boser and colleagues presented a method that enhances training data margin. The SVM finds the optimum hyperplane to divide data from two classes with the highest margin and classifies data linearly or nonlinearly based on features \cite{boser1992training,sarlak2011svm}. 
    \item $K$-Nearest Neighbor (KNN): The $K$-nearest neighbor's method was made by Evelyn Fix and Joseph Hodges. The main idea behind the classification algorithm is that an object is in the same class as its $K$ most similar neighbors if most of those neighbors are in the same class \cite{cover1967nearest}. 

    \item  Random Forests (RF): Random forest is an ensemble learning method for classification, regression, and other problems that builds many decision trees during training and output \cite{breiman2001random}. Forests combine decision tree techniques, utilize several trees, and improve a single random tree.
    \item Naïve Bayes (NB): NB is the simplest Bayesian network classifier \cite{langley1992analysis}. NB classifiers need linear parameters for the number of variables (features/predictors) in a learning problem. Classifiers are usually built using NB assigning class. 
    \item Decision Tree (DT): The decision tree model is very common among classifiers \cite{breiman1984classification}. DTs are well-liked due to their usefulness and simplicity. Additionally, rules could be retrieved conveniently from DTs. The classification process is performed using classification trees, which employ a top-down methodology by dividing the input training data into smaller branches until a branch is identified that exhibits the most pertinent label. The DT consists of a central ``root" node, multiple ``nodes," ``branches," and ``leaves," each representing class labels. 
\end{enumerate}

\subsubsection{Neural network-based methods} 
\begin{enumerate}[i.]
    \item Deep Neural Network (DNN): The structure of a DNN is made up of $n$ parameters, which are passed through a network with $N$ layers to get the results and it is represented as:
    
    \begin{equation}
    h_1= f_1 (W_1^T \, X + b_1)
    \end{equation}
    
    The input value is $X$, the first layer's weight is $W_1^T$, and the first layer's bias is $b_1$. The first layer's result is $h_1$, which is sent to the second layer. In the case of multiple artificial neural networks, their primary learning process involves assessing a single epoch and subsequently adjusting the weight and bias values based on the error, to reduce the error within each layer. This approach is commonly referred to as error backpropagation. Neurons, weights, biases, and functions are the same components of all neural networks, regardless of their shape. These components could be trained just like any other machine-learning algorithm because they function similarly to the human brain. The neurons that are located in the layer below are connected to every neuron in this layer. It has the most straightforward structure and serves as a crucial connection between all of the neurons in the layers above and below. If all of the neurons in all of the layers are fully connected (FC), the neural network is referred to as a DNN. Figure \ref{fig:FIG10}  illustrates the DNN's structural layout. Supervised deep learning approaches enable the construction of predictive algorithms for intricate decision-making scenarios, such as the assessment of an object's condition.
    \begin{figure*}[htp]
    \centering
    \includegraphics[width=12cm]{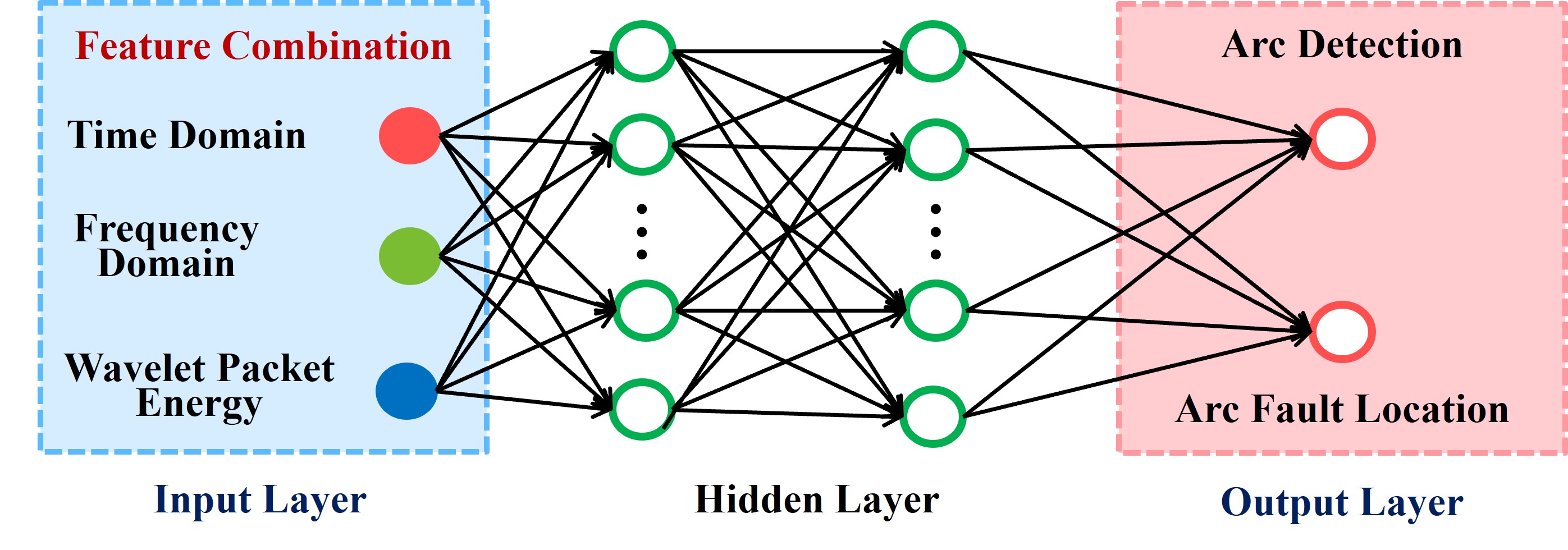}
    \caption{The architecture of DNN employed by \cite{jiang2021series} is utilized for the purpose of identifying and locating arc faults.}
    \label{fig:FIG10}
\end{figure*}
    In the domain of fault detection, this condition is determined by the outcome of a prediction of a potential failure or the identification of an already-existent failure in a particular entity. In essence, a fault detection system that utilizes deep learning techniques is employed in machine vision scenarios and involves extensive data processing to detect and/or anticipate errors. 
{    \item  Recurrent Neural Network (RNN): The RNN is commonly employed to characterize sequential data. Currently, there are numerous effective applications of this technology in various fields such as speech recognition, natural language processing, medical diagnosis, fault diagnosis, and other domains. To gain a comprehensive understanding of the training process of RNNs, it is crucial to highlight the significant distinctions between feed-forward networks and RNNs. Feed-forward networks are classified as stateless models, as they do not possess dynamic states. This is because, throughout the training process, the data is handled as static and does not involve any temporal dependencies. Feed-forward neural networks are used to depict the correlation between variables in each observation at a certain moment, without taking into account the history. Therefore, significant links could be lost in certain cases as a result of neglecting the influence of dynamic correlations between past and present input and output data. The notion of ``stateful" models was formulated to effectively exploit the sequential dependencies of data pertaining to previous inputs by persistently retaining the information in a ``memory cell" over a period of time \cite{siegel2018real}.}
    \item Long Short-Term Memory (LSTM):
    LSTM is part of the recurrent neural network (RNN) method. There are three gates in an LSTM unit: a ``forget" gate, an ``input" gate, and an ``output" gate. Through selective forgetting and remembering mechanisms, these gate structures could get good feedback on the right information, which helps the network get a better approximation of complex nonlinear functions that change over time. In LSTM, the short-term memory is handled separately from the long-term memory. The output of an LSTM neuron is represented by the following equation:

    \begin{equation}
     y_t = (\sigma \, (h_{t-1} \, x_t + W_0 \, h_{t-1} + C_t \; C_{t-1} + b_0)) \; \tanh (c_t)
    \end{equation}

    where $\sigma$ is a variable that controls how much the weight and bias values change based on the data received in one repetition (the learning rate). $h_{t-1}$ and $h_t$ are the current and past states of short-term memory, respectively. $C_{t-1}$ and $C_t$ are the current and past states of long-term memory, respectively. $W_0$ and $b_o$ are the current cell's weight and bias, respectively. The data that come from another LSTM cell are represented by $x_t$. The outgoing data, which is sent to another LSTM cell, is shown by the variable $y_t$. The foundational architecture of LSTM is shown in Figure \ref{fig:FIG11}. 
  \begin{figure*}[htp]
    \centering
    \includegraphics[width=12cm]{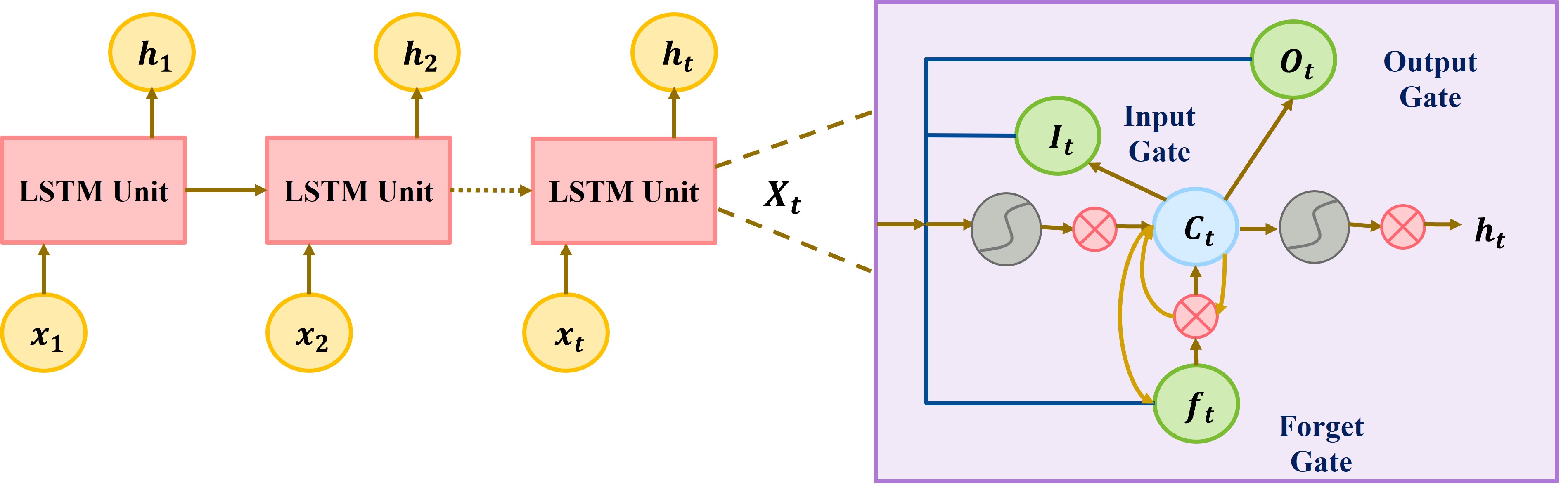}
    \caption{The foundational architecture of LSTM.}
    \label{fig:FIG11}
\end{figure*}

    \item Gate Recurrent Unit (GRU): GRU is another RNN algorithm. In GRU, long-term and short-term memory are merged, unlike LSTM. GRU has two main gates: update and reset. The update gate controls past state information flows into the present state. The reset gate controls the ignored status information at the last minute. The equation below represents the output of a GRU neuron:

    \begin{equation}
    h_t = z_t \, \hat{h}_t + (1 - z_t) \, h_{t-1}
    \end{equation}

    Memory states at the past and current times are $h_{t-1}$ and $h_t$, respectively. $z_t$, $\hat{h}_t$ represent the update gate and candidate activation vectors.

    \item Gated Recurrent Neural Networks (GRNNs): have demonstrated their efficacy in various domains that involve sequential or temporal datasets, including but not limited to natural language processing, music synthesis, speech recognition, machine translation, and classification tasks \cite{dey2017gate,james2017intelligent}.

    \item Convolutional Neural Network (CNN): The CNN architecture consists of a number of interconnected network layers. The typical CNN architecture, as described in reference \cite{abdeljaber20181}, typically comprises an initial input layer, several convolutional and pooling layers, one or two fully connected layers, and a final output layer.
    
\begin{itemize}
    \item The primary function of the input layer is to receive input data within the model. {The quantity of neurons within this particular layer corresponds to the number of input values}, specifically denoting the number of pixels in the context of a picture.
    \item The output received from the input layer is subsequently transmitted to the hidden layer. The presence of numerous concealed layers is contingent upon the specific model and magnitude of the data. Typically, the number of neurons in each hidden layer tends to exceed the number of features. The calculation of the output for each layer entails performing {convolutional layers that use filters (or kernels) to scan over the input data (e.g., an image), which could be adjusted through learning.} Subsequently, the learnable biases are incorporated, and an activation function is applied to create nonlinearity into the neural network.
    \item The output layer is responsible for receiving the data from the hidden layer and applying a logistic function, such as sigmoid or softmax, to convert the output of each class into a probability score representing the likelihood of that class. The activation functions commonly utilized in diverse machine learning models include rectified linear unit (ReLU), hyperbolic tangent ($\tanh$), and sigmoid whose curves are shown in Figure \ref{fig:FIG12}.

\begin{figure}[t]
    \centering
    \includegraphics[width=10cm]{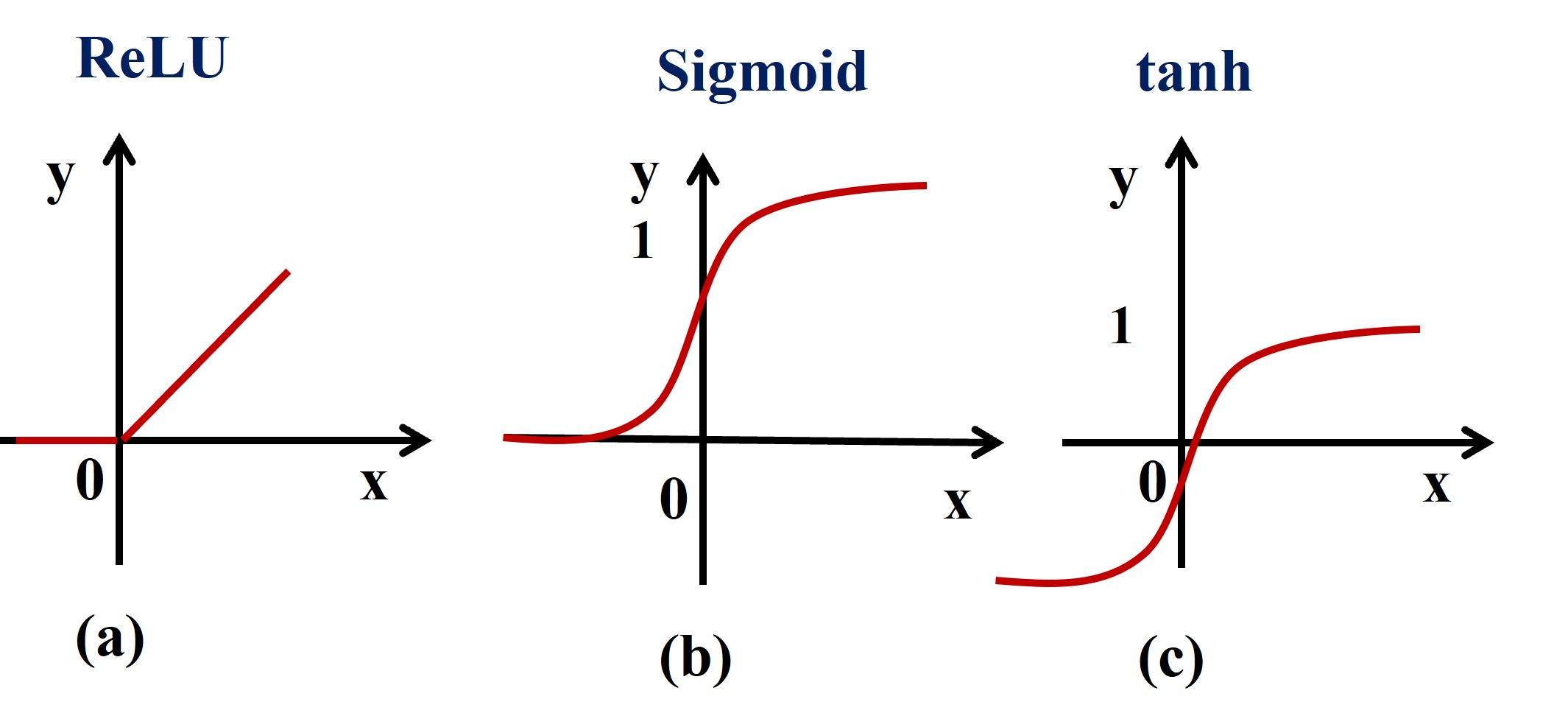}
    \caption{Activation curve.}
    \label{fig:FIG12}
\end{figure}
\end{itemize}
\end{enumerate}
\begin{table}
\caption{AI Techniques Used For Arc Fault Detection}
\centering
\renewcommand{\arraystretch}{1.5}
\begin{tabular}{>{\centering}m{16em} c c}
\toprule
  \multicolumn{1}{>{\raggedright}m{16em}}{AI methods} &
  \multicolumn{1}{>{\centering}m{6em}}{References} & \\
  \midrule
   \multicolumn{1}{>{\raggedright}m{16em}}{Support Vector Machines (SVM)} &  
   \multicolumn{1}{>{\centering}m{6em}}{\cite{miao2020dc,wang2021arc,9431207,shen2021wavelet,dang2021series}}   \\ 
  \multicolumn{1}{>{\raggedright}m{16em}}{$K$-Nearest Neighbor (KNN)} & 
  \multicolumn{1}{>{\centering}m{6em}}{\cite{dang2021series}} \\
  \multicolumn{1}{>{\raggedright}m{16em}}{Random Forests (RF)} & 
   \multicolumn{1}{>{\centering}m{6em}}{\cite{dang2021series}, \cite{jiang2021series}}\\
  \multicolumn{1}{>{\raggedright}m{16em}}{Naïve Bayes (NB)} &  
   \multicolumn{1}{>{\centering}m{6em}}{\cite{dang2021series}}\\
  \multicolumn{1}{>{\raggedright}m{16em}}{Decision Tree (DT)} & 
   \multicolumn{1}{>{\centering}m{6em}}{\cite{dang2021series}}\\
  \multicolumn{1}{>{\raggedright}m{16em}}{Deep Neural Network (DNN)} &  
  \multicolumn{1}{>{\centering}m{6em}}{\cite{yu2019identification, zhang2022series, gong2023series}}\\
  \multicolumn{1}{>{\raggedright}m{16em}}{Long Short-Term Memory (LSTM)} &  
  \multicolumn{1}{>{\centering}m{6em}}{\cite{dang2021series},\cite{sung2022tl}}\\
  \multicolumn{1}{>{\raggedright}m{16em}}{Gate Recurrent Unit (GRU)} & 
   \multicolumn{1}{>{\centering}m{6em}}{\cite{dang2021series}}\\
  \multicolumn{1}{>{\raggedright}m{16em}}{Gated Recurrent Neural Networks (GRNNs)} & 
   \multicolumn{1}{>{\centering}m{6em}}{\cite{wang2023dc}} \\
  \multicolumn{1}{>{\raggedright}m{16em}}{Recurrent Neural Network (RNN)} & 
   \multicolumn{1}{>{\centering}m{6em}}{\cite{9206574}} \\
  \multicolumn{1}{>{\raggedright}m{16em}}{Convolutional Neural Network (CNN)} & 
   \multicolumn{1}{>{\centering}m{6em}}{\cite{wang2023line,gao2023fault,s23177646, yang2019novel,jiang2023ac}}\\
\bottomrule
\end{tabular}
\label{tab:AI_methods}
\end{table}

\begin{figure}[t]
    \centering
    \includegraphics[width=9cm]{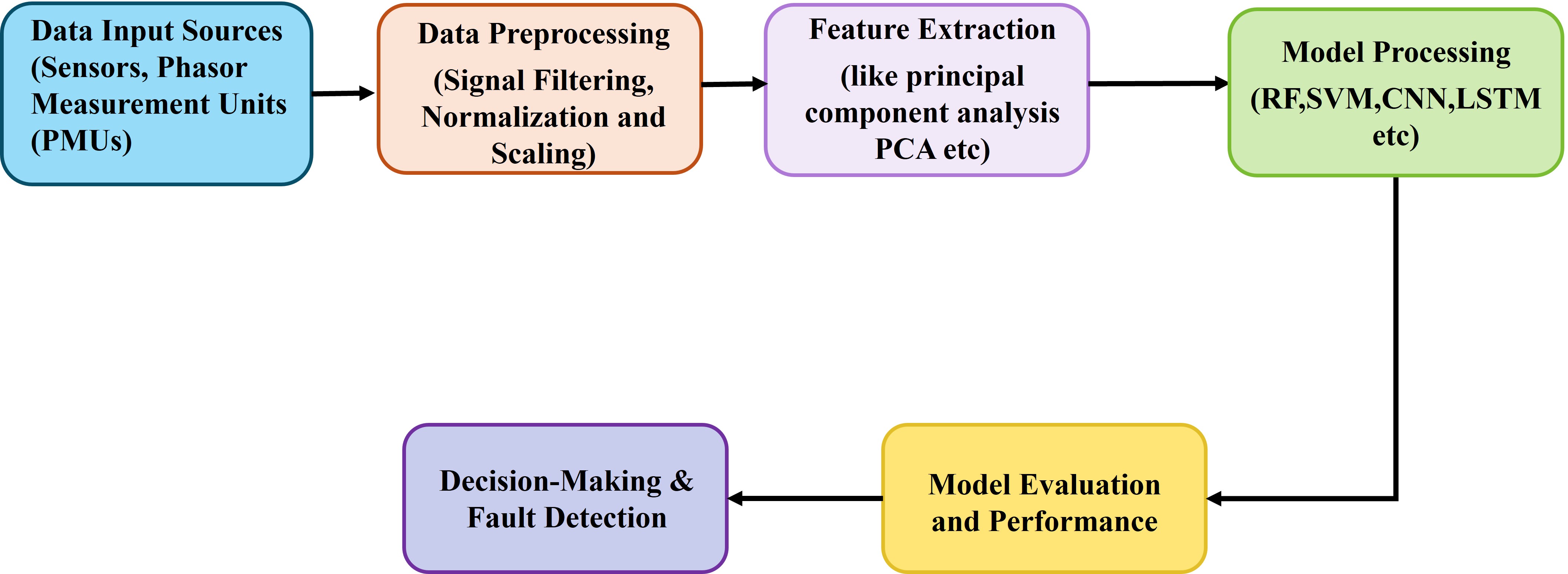}
    \caption{{Outline of arc fault detection system architecture.}}
    \label{fig:AFDSA}
\end{figure}

{Figure \ref{fig:AFDSA} illustrates an AI-driven arc fault detection system for electrical networks. Data from sensors and PMUs is subjected to preprocessing and feature extraction to emphasize fault signs. An AI model analyses these properties to detect arc faults, activating warnings or automatic responses upon detection. This architecture guarantees effective, instantaneous fault detection and response.}

\section{Challenges and Opportunities}
\label{section:challenges_opportunities}

The incidence of electrical safety accidents resulting from arc faults has significantly contributed to the overall number of incidents and has exhibited a progressive upward trend. The aforementioned phenomenon presents a growing threat to both human lives and material possessions. 
The research of arc fault detection technology has gained significant relevance and widespread societal demand due to its crucial application background in addressing and mitigating electrical mishaps caused by arc faults in residential buildings. The topic exhibits a strong correlation with the safety of residential electricity, as well as the overall safety of individuals and their belongings. Both local and international investigations have shown favorable outcomes in the domain of arc fault detection technology, as well as innovation, advancement, and standardization of application products.

For accurately identifying the complexity, trends, and breakthroughs in the context of arc fault identification, a thorough understanding of the literature is essential. In addition to enhancing knowledge, a thorough analysis of the existing literature could help reveal gaps and inequalities that could be used to lay the groundwork for future research projects. A comparison from the recent research articles has been done in the subsequent subsections.

\subsection{Challenges of Traditional Arc Fault Detection Approaches}
Extensive research has been conducted in recent decades to investigate the concerns and challenges associated with arc fault detection. Detecting arc fault features involves defining criteria for physical values such as the low-frequency harmonic value of the chirp zeta transform or the thermal features generated by the arc. Nevertheless, monitoring the physical quantity of all circuit components in real time poses a challenge in power distribution systems that involve complicated loads. In recent years, the academic literature pertaining to arc fault detection has mostly focused on two key areas. One perspective to consider is the selection of suitable eigenvalues in order to examine the underlying physical attributes of circuit faults. In contrast, the investigation of data processing techniques is conducted with the aim of attaining arc fault detection. In \cite{ji2018optimal}, a bandpass filter was constructed to isolate arc signals within the frequency range of 2.4 to 39 kHz. Subsequently, the effectiveness of the db13 wavelet in analyzing the arc characteristics was confirmed through the application of the discrete wavelet transform method. 

However, it was determined through experimental analysis that these methodologies have limitations in accurately capturing the unique characteristics of different loads. This is due to the dynamic nature of electrical appliances, which are always evolving and introducing creative features. This article in \cite{9957132} introduces a detection methodology that employs signal-type enumeration and a zoom circular convolution (CC) technique (ZCC). The signal-type enumeration technique enhances the detection method's ability to generalize under unfamiliar conditions. The many components of current with impulsive characteristics could be categorized into stable, periodically impulsive, nonperiodically impulsive, and hybrid signals through a process of enumeration. Given the constraints of CC, the ZCC technique is introduced to extract discernible features and reduce the computational cost associated with CC. Additionally, the signal function is partially recreated to enhance the performance of CC.
The primary challenge encountered during the creation of AFCI is the selection of one or more appropriate criteria. Various approaches for detecting arcing faults have been put forward in academic literature, with spectrum analysis being the most commonly employed detection technique. The fault detection technique proposed in \cite{JOVANOVIC201611} utilizes the computation of the energy density spectrum and is designed to operate within the framework of a pre-existing shunt active power filter architecture.
The methodology outlined in \cite{mukherjee2015method} examines the power spectrum density of electromagnetic radiations.

A comprehensive understanding of the causal connections between problems and symptoms exhibited by transformers is required. Therefore, many technicians encounter difficulties while attempting to implement prior knowledge approaches.
In general, data-driven procedures tend to exhibit higher levels of accuracy compared to knowledge-driven ones. The accuracy of physical models is typically inferior to that of data-driven models. If sensors are put in a manner inconsistent with the specified requirements, extensive changes will invariably be required. Typically, it is advisable to have a qualified professional perform the necessary modifications to equations and parameters.

Nevertheless, the conventional approaches encounter challenges such as the complexity of setting thresholds, extensive computational demands, and limited capacity for generalization when employing FFT, DWT, and other advanced algorithms. To expedite the process of computation, it is necessary to employ methods that yield rapid results. To achieve real-time detection, the time-frequency information of the circuit is considered the primary selection for fault feature value. A wavelet compression reconstruction-based data enhancement technique addresses neural network overfitting caused by few samples. Mathematical models pertaining to arc faults have been created based on data from experiments. Models do not fully reflect the arc's exterior properties and are helpful for theoretical investigations only. 

Moreover, there have been advancements in arc fault detection approaches that rely on the identification of specific features associated with arc faults, including strong luminosity, elevated temperatures, significant distortion noise, and enhanced electromagnetic radiation \cite{kim2021dc}. However, a significant limitation of these devices is their inherent inability to accurately determine the precise locations of arc faults. In \cite{koziy2013low}, the drop mode change signals of voltage and current were detected using wavelet transform. The root mean square of voltage and current was then retrieved as the indicators for analysis. The techniques employed in this study are founded upon the wavelet transform, a versatile mathematical tool that may be effectively utilized across various load types, hence enhancing the accuracy and dependability of detection mechanisms.
Also, the aforementioned indicator threshold values are established in an arbitrary manner, hence lacking both flexibility and dependability in their determination. The methods for detecting series arc faults, which rely on statistical analysis and signal processing, need a substantial level of experimental expertise. These methods are most appropriate for systems with relatively small sample sizes and a limited range of load types. On the other hand, the utilization of neural network-based series arc defect detection methods is more suited for extensive systems that encompass a wide range of load types.
In addition, it has been observed that certain commercial AFCIs tend to either fail to activate when required or activate erroneously. A previously conducted research study \cite{su2010assessment} reported an approximate 50\% accuracy rate in detecting arc faults. Consequently, the correct detection of all arc faults in circuits remains challenging, and there is a need for further refining of some detection methods, particularly for circuits operating at 220-240 Volt.

\subsection{Promising Opportunities for AI/ML-Based Techniques and Hybrid Approaches}

There is a growing interest in the utilization of machine learning/artificial intelligence-based approaches for arc fault detection applications, primarily driven by their efficiency and accuracy. Within the field of electrical system protection, researchers have observed an important shift focused on the incorporation of data analytics. The results of these attempts are remarkable. Researchers are currently directing their efforts toward leveraging data analytics in order to optimize the intelligence and effectiveness of protection mechanisms within electrical systems. The convergence of intelligent technology and data-driven insights hold great potential for enhancing the effectiveness and dependability of electrical system protection \cite{MP_USpatent,MP_WOpatent,dwivedi2023identification,dwivedi2022evaluation,Dwivedi_2023}.

Recent research {\cite{wang2023line,s23177646,le2020series,10054597,jiang2023ac}} has successfully demonstrated the effectiveness of AI-based approaches in detecting series arc faults, yielding excellent outcomes. { Several research studies have been carried out to examine the detection of arc faults. These investigations are classified according to variations in voltage levels (high and low), current magnitudes (high and low), applicative fields (residential and industrial), and the differentiation between DC and AC.} Upon reviewing the prior research, it has been noted that the majority of studies primarily focus on utilizing arc data as current. However, only one study has incorporated both voltage and current as arc data, while another study solely employed voltage as arc data as shown in Table \ref{tab:compare_dataset}. 

\begin{table}
\centering
\renewcommand{\arraystretch}{1.5}
\caption{Comparison of different types of data for arc faults.}
\begin{tabular}{>{\centering}m{2em} c c}
\toprule
 \multicolumn{1}{>{\centering}m{5em}}{Data Type} & 
    \multicolumn{1}{>{\centering}m{5em}}{Data Source} & 
    \multicolumn{1}{>{\centering}m{14em}}{References} \\
\midrule
 \multicolumn{1}{>{\centering}m{5em}} {Current} & 
\multicolumn{1}{>{\centering}m{5em}}{Field \& Simulation} & 
\multicolumn{1}{>{\centering}m{14em}}{\cite{wang2021arcnet,chu2020series,yu2019identification,miao2020dc,zhang2022series,wang2021arc,wang2022novel,10054597,9206574,wang2023line,9431207,gao2023fault,jiang2023ac,calderon2019kalman,jiang2023non,8354947,luan2022arc,s23177646,yang2019novel,zhao2022series,jiang2021series,li2023low,sung2022tl,shen2021wavelet,MENG2023109286,han2020series,le2020series,chabert2023transformer,wang2023dc,dang2021series,gong2023series}}\\ 
 \multicolumn{1}{>{\centering}m{5em}} {Voltage and Current }  & 
\multicolumn{1}{>{\centering}m{5em}}{Field} & 
\multicolumn{1}{>{\centering}m{14em}}{\cite{han2021recognition}}\\
 \multicolumn{1}{>{\centering}m{5em}} {Voltage} & 
\multicolumn{1}{>{\centering}m{5em}}{Field \& Simulation}  & 
\multicolumn{1}{>{\centering}m{14em}}{\cite{lala2020detection}}\\
\bottomrule
\end{tabular}\\
\label{tab:compare_dataset}
\end{table}

According to the literature, researchers commonly employ both linear and nonlinear loads in their studies. Some researchers exclusively focus on nonlinear loads, while others solely investigate linear loads.

The impact of a series arc on line current varies depending on the load, posing challenges in reliably extracting arc fault characteristics that are applicable to all loads based on the line current signal \cite{han2020series}. As current characteristics fluctuate with load, the generalization performance of the determined detection method would be compromised when applied to undetermined loads. Although current characteristics vary among different categories of loads, even identical types of loads manifest distinct current characteristics. In light of the current characteristics of diverse loads, it is imperative to establish an adequate quantity of scenarios encompassing various loads and choose an adequate number of features to identify faults \cite{jiang2023ac}. Hence, the literature has extensively examined arc fault detection methods that have various types of loads, which are discussed below.

\subsubsection{Combined Linear and Nonlinear Loads}
The procedure of detecting arc faults in electrical systems encompasses the identification of potentially dangerous electrical arcs within a system that has a combination of linear and nonlinear loads. Linear loads provide a steady and predictable current draw pattern that closely resembles a sinusoidal waveform. Nonlinear loads, which are prevalent in contemporary electronic devices, exhibit a more intricate and non-sinusoidal current draw pattern. The identification of arcs under complex load circumstances poses a substantial obstacle. Linear loads have the capacity to induce anticipated disruptions in the waveform of the current, whilst nonlinear loads possess the capability to introduce unforeseeable fluctuations. Detection methods must take into consideration the wide range of diversity present. Examples of linear and nonlinear loads are incandescent lights, resistive heaters, motor-type loads, power electronic loads, microwave ovens, and similar electrical appliances \cite{wang2022novel,jiang2023ac,10054597}.

An all-encompassing strategy could involve the utilization of specialized sensors and monitoring equipment for the continuous analysis of electrical waveforms. Machine learning algorithms have the capability to undergo training in order to distinguish between typical load fluctuations and typical patterns that signify the presence of an arc. To achieve accurate identification, the model must be capable of adapting the diverse properties exhibited by both types of loads. The implementation of this integrated technique improves safety by effectively detecting potential arc faults within a system that encompasses a variety of load types.

The primary focus of studies in studying arc faults in both linear and nonlinear loads has been largely centered on scenarios employing AC. In this particular field, a significant portion of research has focused exclusively on low-current conditions. On the other hand, a certain group of researchers has focused their efforts on examining DC situations.
Table \ref{tab:both_linear_and_non_linear} presents a comparison of the approaches employed by different researchers in analyzing both linear and nonlinear loads.

\begin{itemize} 
    \item \textbf{AC Arc Fault Detection}: ArcNet convolutional neural network-based arc identification model is proposed in \cite{wang2021arcnet}. They detected AC series arc fault. The database includes two of the most frequent forms of arcs: those resulting from loose cable connections and those arising from insulation failure, both of which are generated during testing. According to test findings, ArcNet could recognize arcs with the greatest precision accuracy of 99.47\% at a sampling rate of 10 kHz while using a database of raw current. The high-frequency coupling convolution neural network (HCCNN) approach for the detection of a series arc fault is presented in \cite{chu2020series}. This sensor collects high-frequency feature signals from different types of loads in a series arc and regular operation states. The temporal-domain sequence converts the signal to two-dimensional feature grayscale images. These photos are used to train and assess a three-layer convolution neural network using a number of common loads' arc conditions and normal conditions data sets. It detects the load type and series arc simultaneously and accurately. DWT in conjunction with a DNN is proposed in \cite{yu2019identification}. An experimental bed was developed to take the current signals under normal and arcing states. The current properties of arc faults were analyzed, and afterward, wavelet coefficient sequences were utilized to construct the input matrix for a deep neural network.{A new hybrid methodology enhances arc fault identification by integrating salp swarm optimization with variational mode decomposition to boost feature extraction and dataset development. Employing ReliefF in conjunction with minimum redundancy maximum relevance (mRMR) algorithm mitigates feature redundancy, whereas a random forest classifier facilitates rapid and precise fault diagnosis, with over 95\% detection accuracy in noisy conditions \cite{10399639}.
    A hybrid combination of principal component analysis and support vector machine (PCA-SVM) is proposed for series arc detection \cite{jiang2019series}.} Accuracy levels for load identification and series arc identification are 99.1\% and 99.3\%, respectively. To examine the properties of DC arc faults, a DC arc generator has been designed and is depicted in Figure \ref{fig:FIG13}.

\begin{figure}[t]
    \centering
    \includegraphics[width=9cm]{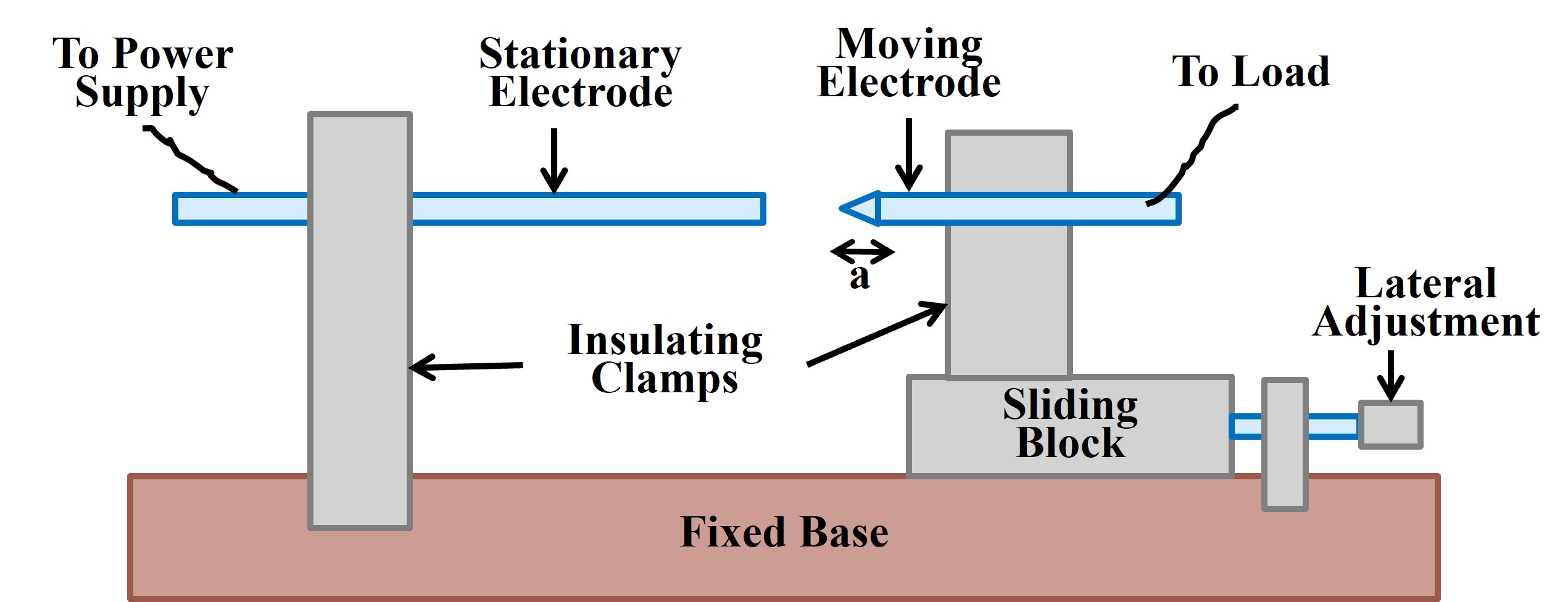}
    \caption{Laboratory arc generator setup used in laboratories for experimental purposes.}
    \label{fig:FIG13}
\end{figure}
A residual network (ResNet)-based arc fault detection model and data enhancement method are presented in \cite{zhang2022series}.  An enhanced Mel-Frequency Cepstral Coefficients (MFCC) for preprocessing and a neural network model named ARCMFCC is proposed for identifying arcs to attain the highest level of accuracy in detecting arcs that is 99.34\% \cite{wang2022novel}.  A novel approach was proposed for detecting series arc faults and selecting phases, utilizing single-phase current \cite{9431207}. The integration of the Fractional Fourier Transform (FRFT) and the 2-level block Singular Value Decomposition (SVD) is highly effective in extracting arc fault detection and achieving phase selection. In \cite{gao2023fault} outlines a way for finding arc fault that combines the residual network framework and the channel attention technique. This research presents a novel module called one-dimensional depth separable convolution (1D-DS) that tries to decrease the number of variables in a network model and lower the inference time for individual prediction samples. The accuracy of their model was found to be 98.07\%. In \cite{jiang2023ac} presents a novel approach for the detection of series arc defects utilizing a specific mechanism, a high-frequency (HF) RLC AM and a one-dimensional convolutional neural network (1DCNN). Better generalization and 99.33\% average detection accuracy are achieved with the provided method compared to the 1DCNN method using limited real-current data. The architecture of the presented 1DCNN is depicted in Figure \ref{fig:FIG14}.

    \begin{figure}[t]
    \centering
    \includegraphics[width=9cm]{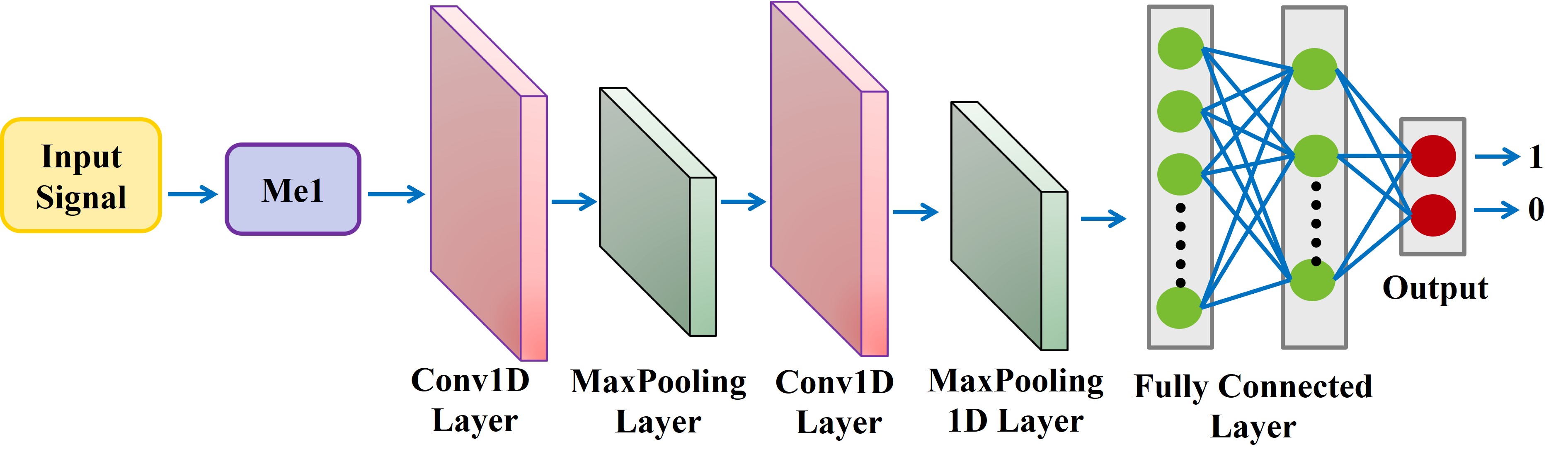}
    \caption{Structure of the proposed 1DCNN \cite{mohammed2023detecting}.}
    \label{fig:FIG14}
\end{figure}

In \cite{jiang2023ac} describes a technique for detecting series arc defects utilizing an HF RLC AM and a 1DCNN. Better generalization and 99.33\% average detection accuracy are achieved with the provided method compared to the 1DCNN method using limited real-current data. The research study in \cite{calderon2019kalman} introduces a novel approach for identifying series arcing problems inside AC residential electrical systems. The method under consideration is founded on the utilization of a Kalman filter for the purpose of detecting fault indications. Additionally, a decision block is incorporated into the algorithm to validate the existence of a series arc fault and then trigger a tripping signal. The estimation of the current at one end of the power line is conducted using a model that incorporates two steady-state variables, namely X1 and X2. The input parameters for failure symptom detection in a Fuzzy logic processor include residuals and the third-order difference of state X2.
Additionally, the fault indications are analyzed by a detecting logic block, which verifies the existence of an electrical arcing problem.
In \cite{jiang2023non}, the Mann-Kendall Test is used to compare two nearby cycles' current waveforms rather than directly using them. The challenge of arc localization is converted into an optimization problem, wherein a genetic algorithm (GA) is employed to maximize the findings of differential current decomposition and accurately determine the location of the arc fault. In order to identify residential AC series arc faults, the research in \cite{8354947} suggests a sparse representation and fully connected neural network (SRFCNN) approach. The SRFCNN approach employs neural networks (NN) to carry out adaptive feature learning and classification while capturing the characteristics of conventional signals using sparse coding. A preprocessing layer, a sparse representation layer, and a decision layer make up the methodology's intricate organizational structure. The following main steps make up the processing process. Prior to selecting a fixed vocabulary, a collection of customized bases is first constructed via dictionary learning to act as accurate feature descriptors. Accordingly, a set of sparse coefficients is created from the pretreatment input samples, which are subsequently fed into the NN for identification. 
In \cite{luan2022arc} methodology for the detection and identification of arc faults in low-voltage users at the service entry. The approach that has been suggested utilizes supervised, non-intrusive current disaggregation. The suggested methodology involves the utilization of a Non-Intrusive Load Monitoring (NILM) module to detect and identify the appliances that are currently in operation within a domestic dwelling. Additionally, an arc fault detection module is employed to examine the simultaneous total current signal for the presence of an arc fault. If an arc fault is identified, a module for arc current disaggregation will employ total current harmonic disaggregation to identify the most suitable arc fault waveform associated with one or many individual appliances. The classification technique known as NILM could be categorized as either supervised or unsupervised, based on the type of data utilized during offline training to provide prior class information for subsequent prediction\cite{wong2013recent}.

In \cite{s23177646}, a method for arc defect identification is based on ensemble learning and decision-level feature fusion. The effectiveness of the multimodal feature fusion technique in detecting arc faults surpasses that of using single-mode features alone. In a study using 1000 randomly selected samples, the average detection rate achieved was 98.87\%, with the highest detection rate of 99.30\%. The present study introduces a novel methodology referred to as the temporal domain visualization convolutional neural network (TDV-CNN) in the manuscript \cite{yang2019novel}. The current transformer and high-speed data collecting system are employed for the purpose of gathering the current measurements pertaining to a succession of arc faults. Subsequently, the acquired signal undergoes digital filtering and is subsequently transformed into a grayscale image in chronological order prior to being inputted into the TDV-CNN. The experimental findings validate that the classification accuracy could attain a level of 98.7\% or perhaps exceed it with optimal parameter adjustments. The implementation of a detection methodology that solely depends on the zero-current time proportional coefficient or the normalized mean square error coefficient leads to the presence of both false alarms and missed alarms when exposed to different types of loads, such as workstations and fluorescent lighting. The approach employed by the researchers in \cite{zhao2022series} offers a distinct benefit as it incorporates the fusion of many features associated with arc faults rather than merely enumerating these features. They have used fuzzy logic for arc identification. Consequently, this approach exhibits a notable level of accuracy in the detection of arc faults.

The non-stationary and nonlinear nature of low-voltage series arc fault currents needs careful consideration of fault features and the adaptation of the detection algorithm to get accurate detection results. 
 In light of the aforementioned concerns, a novel approach called ECMC is introduced by authors in \cite{li2023low}, which integrates Euclidean distance, classifier criterion, max-relevance min redundancy(mRMR), and clustering index as a feature selection method having optimal feature subset Fb. This approach has the capability to eliminate features with weak correlations and duplicate information in the domains of time, frequency, and time-frequency. The feature data set created by ECMC is able to comprehensively capture the distinctive attributes of arc faults. The suggested approach introduces a stochastic configuration network (SCN) that utilizes variational Bayesian (VB) optimization, referred to as VB-SCN. To iteratively improve the output weights and scale functions of the SCN, the suggested method makes use of VB. The utilization of VB-SCN has the potential to enhance the overall quality of hidden-layer nodes inside a network, as well as improve its generalization ability. This is achieved by simultaneously ensuring the adaptive learning of current characteristics, ultimately resulting in a faster and more stable arc defect detection process. The flow chart of the ECMC feature selection approach is depicted in Figure \ref{fig:FIG15}.
\begin{figure}
    \centering
    \includegraphics[width=8cm]{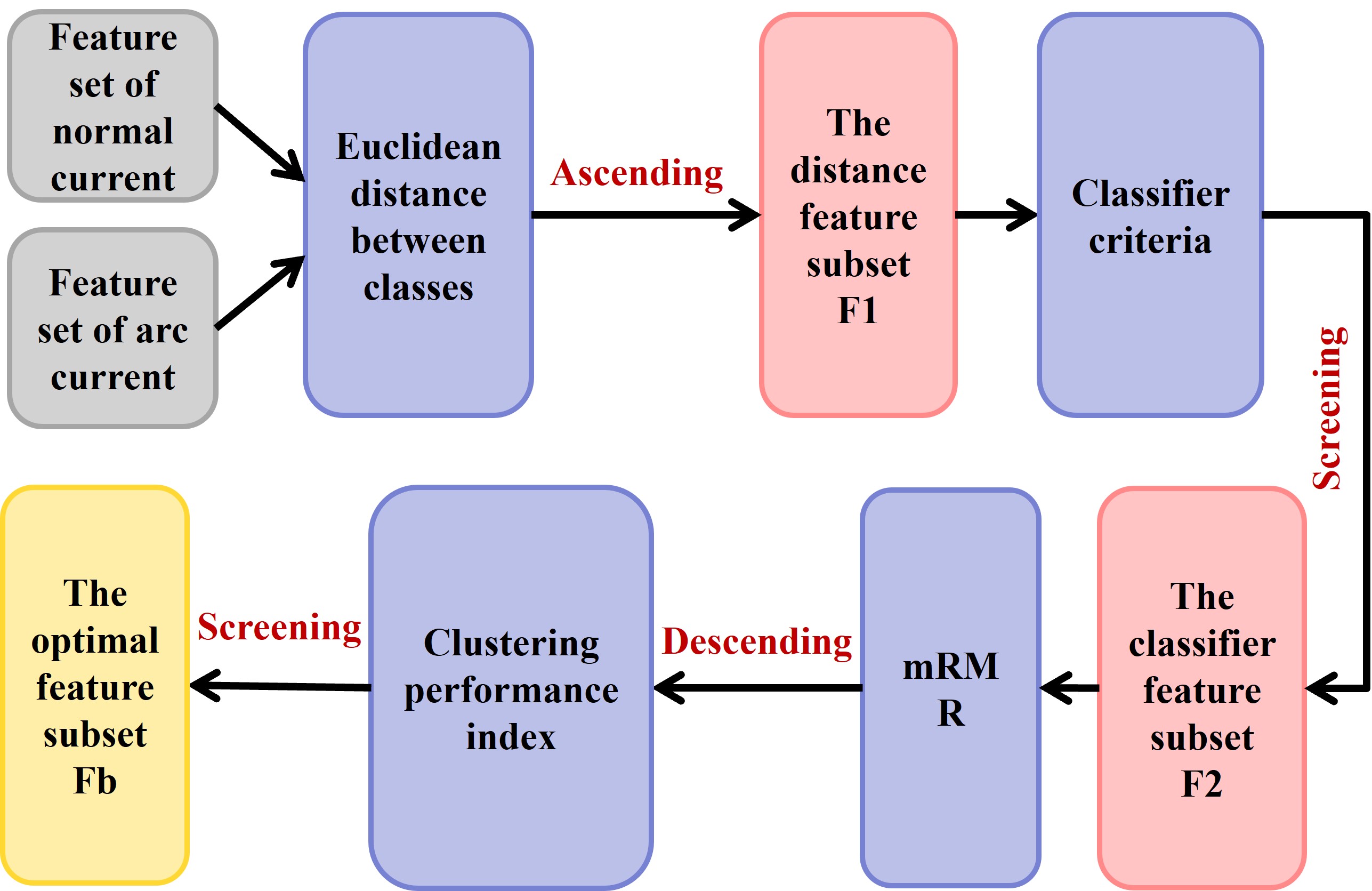}
    \caption{Flow diagram depicting the ECMC feature selection approach \cite{li2023low}.}
    \label{fig:FIG15}
\end{figure}
The study presents in \cite{jiang2021series} a novel technique for detecting and locating series arc faults in multi-load circuit topologies, with a particular focus on branch arc faults and nonlinear power loads. This study examines the current fluctuations resulting from the occurrence of arcing at various locations in paralleling coupled loads. In this study, a feature selection technique utilizing the RF algorithm is employed to identify particular feature sets based on the minimization of Gini impurity. The DNN utilizes a selection of the top ten features that exhibit a strong correlation with the arc. These features are chosen based on their integration and are inputted into the DNN for calculation and training, considering various combinations of loads.
The author of \cite{colo2023intelligent} suggests that the connection with SimulAI enables the seamless processing of input data derived from simulators as well as actual devices, such as sensors and cameras. Furthermore, the inclusion of modularization provides the opportunity for interaction with other elements in a closed-loop manner. A study in \cite{shen2021wavelet} proposed a data-driven approach for identifying low-energy arc-faults. However, this approach was limited to AC circuits and did not differentiate between low-energy arc-faults and stable arc-faults. Additionally, the trained model was not evaluated in a real-world setting due to the intricate nature of the preprocessing steps required for the SVM classifier\cite{shen2021wavelet}.

This study in \cite{MENG2023109286} involves the acquisition of arc fault signals from a multi-branch experimental platform that has been developed with masking loads in accordance with the specifications outlined in IEC 62606. The inclusion of adaptive noise in the ICEEMDAN algorithm has been determined to be an appropriate approach for extracting a greater amount of comprehensive arc fault information. The IMF is chosen, and its variance is computed to create detection variables that could differentiate between the occurrence of an arc fault and the normal operating condition. On average, the capability to gather arc fault information is enhanced by a factor of 8.47 when comparing it to the wavelet transform. The utilization of Boruta and LightGBM algorithms is employed to enhance the effectiveness of the algorithm and streamline its structure by reducing the dimensionality of the features. The presented approach for arc fault detection effectively and efficiently detects multi-branch arc faults using a well-trained LightGBM model. The Raspberry Pi platform has been demonstrated to achieve an identification accuracy of 97.06\% and a detection time of less than 300 ms.

This study in \cite{han2020series} examines the attributes of several load categories in the presence of a defective arc. Subsequently, a technique for detecting series arc faults is introduced, which relies on category identification and artificial neural network techniques. Initially, the loads are classified into two levels depending on the voltage waveforms and current waveforms. Subsequently, quantifiable measurements are chosen by the distinctive characteristics exhibited by various load categories. The data pertaining to indicators is utilized to train an artificial neural network that corresponds to the type of loads. This trained network is subsequently employed in the detection of arcs. By the designated time frame outlined in the detection criteria, each occurrence of an arc is recorded and tallied by incrementing the arc count by one. Ultimately, the cumulative count result is obtained by integrating the individual arc counts. When the number of counts reaches the specified threshold outlined in the norms, it is considered to indicate an arc fault has occurred. Consequently, the protective device promptly interrupts the circuit.
The authors in \cite{chabert2023transformer} suggest a deep learning strategy devoid of descriptions. To solve this time series issue, they modified a sequence-based model known as a Transformer Neural Network (TNN) model. They used a sequence-to-sequence model to reuse the transformer's encoder. A window of electric current with a minimum of one period of the signals is provided as input to the model.
The labels for every single point in the input window are the output. This necessitated coming up with a novel way to classify the signals, therefore creating an automated technique to increase training supervision.
Their model had a 2\% false positive rate and a 96.3\% identification accuracy.
Certain researchers have put forth the idea of utilizing a neural network-based sensor for the purpose of detecting DC arc faults.
Initially, an arc signal acquisition module is developed using electromagnetic induction. This module is designed to automatically adjust the sample frequency and accomplish signal amplification. The maximum achievable sampling frequency is 4 MHz.
 In \cite{dang2021series}, they have two types of scenarios in which test data of a category might be classified. The first scenario occurs when the test data has already undergone training, whereas the second scenario occurs when the test data has not undergone training. The aforementioned scenarios are commonly categorized as enclosed kinds and unenclosed types, respectively. Various artificial intelligence methods, including machine learning and deep learning (DL) algorithms, were employed in the study. The results indicated that the performance of different AI systems in arc identification varied significantly. Machine learning algorithms provide strong performance in the low-frequency region; nonetheless, they necessitate the extraction of relevant features. {Deep learning algorithms typically require a significant quantity of data in order to achieve optimal performance.} The selection of an appropriate technique is of paramount importance in the creation of a dependable and resilient arc fault detection method.
  A novel neural network approach is proposed in \cite{gong2023series}, which incorporates wavelet analysis and feature value decomposition. The data has been divided into five layers using the DB5 wavelet base. The coefficients of the first layer are then extracted in order to generate the Hankel matrix. The initial process of feature extraction and compression has been completed. Upon performing Eigenvalue Decomposition (EVD) on the matrix, the eigenvector is generated, completing the second phase of feature extraction and data compression. The third stage of feature extraction and data reduction is achieved by computing the mean, root mean square, and standard deviation values of the eigenvector. The DNN fault detection model is trained using three input values with a compressed input scale of 1 × 3. This study introduces a deep neural network (DNN) defect detection model that exhibits low complexity and high timeliness. The complicated nature of the algorithm has been significantly decreased in terms of both operating data and the network's architecture, resulting in a training time of around 33 seconds. The time-efficient model detects faults using current half-cycles. An overall identification rate of 98.7\% was achieved.
   \item  \textbf{DC Arc Fault Detection}: In \cite{10054597}, a technique to speed up feature extraction while maintaining accuracy is proposed. First, DC arc datasets are created using a fully automated process. The usual frequency segment of a DC arc is designed to be extracted using a composite bandpass filter. Additionally, a temporal convolution network-based arc detection neural network is presented to extract current waveform properties. These features are processed using principal component analysis to lessen correlation. Finally, the classifier is a single hidden layer neural network. The database has been assembled from various working situations and scenarios. At a sampling rate of 250 kHz, the arc fault detector could measure DC raw current with a test set accuracy of 99.88\%. With an average real-time detection of 0.15 seconds per sample, the model is also implemented on the Jetson Nano.
   In \cite{le2020series} introduces an efficient and adaptable method for detecting series DC arc faults using ensemble machine learning (EML) methods. A buck converter and a boost converter, both serving as constant power loads (CPLs), have been conceived and constructed to investigate various data to train machine learning approaches and arc fault behaviors. The experimental data is used to extract a set of time domain features, which are then analyzed using the feature significance attribute. The characteristics were further processed by an adaptive normalization method to address the issue of false positive categorization resulting from load changes. The suggested methodology involves a two-step algorithm for the identification and detection of arc faults across various load categories. Numerous pre-processing (descriptors) linked to a machine learning model (or deep learning) are used by the majority of detection approaches in the detection literature. The descriptors for these methods must be manually created. 
   Additionally, the authors in \cite{wang2023dc} suggest a novel approach called the Leftover Gated Recurrent Neural Network (LG-RNN) which aims to extract comprehensive features from the contextual information and afterward carry out classification tasks. Figure \ref{fig:FIG17} illustrates the experimental configuration commonly employed by researchers for investigating series arc faults.

\end{itemize}

\begin{table*}
\caption{Comparison of methodologies used for both linear and nonlinear loads.}
\centering
\renewcommand{\arraystretch}{1.5}
\begin{tabular}{c c l c c}
\toprule
  \multicolumn{1}{>{\centering}m{4em}}{Reference} &
  \multicolumn{1}{>{\centering}m{2em}}{Load} &
  \multicolumn{1}{l}{Methodology} &
  \multicolumn{1}{>{\centering}m{4em}}{Accuracy} & \\
  \midrule
\cite{wang2021arcnet}         & 4       & CNN     & 99.47\%  \\ 
\cite{chu2020series}          & 8       &  HCCNN  & 99.20\%  \\ 
\cite{yu2019identification}   & 4       & DWT and DNN & 97.75\% \\ 
\cite{zhang2022series}        & 6       & Residual Network (ResNet), DNN, wavelet compression reconstruction   &  97.69\%  \\ 
\cite{wang2022novel}          & 12       & MFCC and NN &  99.34\%  \\ 
\cite{10054597}               & 2        & Composite bandpass filter (CBF), principal component
analysis (PCA), temporal convolution network (TCN)  & 99.88\%  \\  
\cite{jiang2023ac}            & 3        & 1-DCNN and RLC-based AM & 99.33\%  \\ 
\cite{calderon2019kalman}     & 7        & Kalman filter and fuzzy logic processor    & - \\ 
\cite{jiang2023non}           & 3        & Mann–Kendall test and current decomposition   & 92.86\%  \\ 
\cite{8354947}                & 6        & Sparse representation and fully connected neural network (SRFCNN)  & 94.30\%  \\ 
\cite{luan2022arc}            & 5        & Non-intrusive load monitoring (NILM) and arc current disaggregation & 94.86\%  \\ 
\cite{s23177646}              & 6        & 1-D CNN & 99.30\%  \\ 
\cite{yang2019novel}          & 10       & TDV-CNN & 98.70\%  \\ 
\cite{zhao2022series}         & 7        & Current fluctuation and zero-current features  & - \\ 
\cite{jiang2021series}        & 4        & Random Forest (RF) - DNN  & 97.50\%  \\ 
\cite{li2023low}              & 4        & ECMC and VB-SCN  & 98.25\%  \\ 
\cite{shen2021wavelet}        & 12       & Wavelet-analysis-based singular-value-decomposition (WASVD), SVM   & 95.80 \%  \\ 
\cite{MENG2023109286}         &  9       & ICEEMDAN and LightGBM algorithm & 97.06\%  \\ 
\cite{han2020series}          & 7        & Wavelet transform, genetic algorithm and ANN  & 99.21\%  \\ 
\cite{le2020series}           & 3        & Ensemble Machine Learning (EML)    & 99.80\%  \\ 
\cite{chabert2023transformer} & 2        & Transformer Neural Network (TNN)    & 96.30\%  \\ 
\cite{wang2023dc}             & 4        & LG-RNN   & 98.27\%  \\
\cite{dang2021series}         & 3        & SVM, random forest, k-nearest neighbor, Naive Bayes, decision tree, LSTM, GRU, DNN
    & 95.00\%  \\ 
\cite{gong2023series}         & 10       & Wavelet transform and DNN    & 98.70\%  \\
\bottomrule
\end{tabular}
\label{tab:both_linear_and_non_linear}
\end{table*}

This study in \cite{MENG2023109286} involves the acquisition of arc fault signals from a multi-branch experimental platform that has been developed with masking loads in accordance with the specifications outlined in IEC 62606. The inclusion of adaptive noise in the ICEEMDAN algorithm has been determined to be an appropriate approach for extracting a greater amount of comprehensive arc fault information. The IMF is chosen and its variance is computed to create detection variables that could differentiate between the occurrence of an arc fault and the normal operating condition. On average, the capability to gather arc fault information is enhanced by a factor of 8.47 when comparing it to the wavelet transform. The utilization of Boruta and LightGBM algorithms is employed to enhance the effectiveness of the algorithm and streamline its structure by reducing the dimensionality of the features. The presented approach for arc fault detection effectively and efficiently detects multi-branch arc faults using a well-trained LightGBM model. The Raspberry Pi platform has been demonstrated to achieve an identification accuracy of 97.06\% and a detection time of less than 300 ms.

This study in \cite{han2020series} examines the attributes of several load categories in the presence of a defective arc. Subsequently, a technique for detecting series arc faults is introduced, which relies on category identification and artificial neural network techniques. Initially, the loads are classified into two levels depending on the voltage waveforms and current waveforms. Subsequently, quantifiable measurements are chosen by the distinctive characteristics exhibited by various load categories. The data pertaining to indicators is utilized to train an artificial neural network that corresponds to the type of loads. This trained network is subsequently employed in the detection of arcs. By the designated time frame outlined in the detection criteria, each occurrence of an arc is recorded and tallied by incrementing the arc count by one. Ultimately, the cumulative count result is obtained by integrating the individual arc counts. When the number of counts reaches the specified threshold outlined in the norms, it is considered indicating an arc fault has occurred. Consequently, the protective device promptly interrupts the circuit.

In \cite{le2020series} introduces an efficient and adaptable method for detecting series DC arc faults using ensemble machine learning (EML) methods. A buck converter and a boost converter, both serving as constant power loads (CPLs), have been conceived and constructed to investigate various data to train machine learning approaches and arc fault behaviors. The experimental data is used to extract a set of time domain features, which are then analyzed using the feature significance attribute. The characteristics were further processed by an adaptive normalization method to address the issue of false positive categorization resulting from load changes. The suggested methodology involves a two-step algorithm for the identification and detection of arc faults across various load categories. Numerous pre-processing (descriptors) linked to a machine learning model (or deep learning) are used by the majority of detection approaches in the detection literature. The descriptors for these methods must be manually created.

The authors in \cite{chabert2023transformer} suggest a deep learning strategy devoid of descriptions. To solve this time series issue, they modified a sequence-based model known as a Transformer Neural Network (TNN) model. They used a sequence-to-sequence model to reuse the transformer's encoder. A window of electric current with a minimum of one period of the signals is provided as input to the model.
The labels for every single point in the input window are the output. This necessitated coming up with a novel way to classify the signals, therefore creating an automated technique to increase training supervision.
Their model had a 2\% false positive rate and a 96.3\% identification accuracy.
Certain researchers have put forth the idea of utilizing a neural network-based sensor for the purpose of detecting DC arc faults.
Initially, an arc signal acquisition module is developed using electromagnetic induction. This module is designed to automatically adjust the sample frequency and accomplish signal amplification. The maximum achievable sampling frequency is 4 MHz. Additionally, the authors in \cite{wang2023dc} suggest a novel approach called the Leftover Gated Recurrent Neural Network (LG-RNN) which aims to extract comprehensive features from the contextual information and afterward carry out classification tasks. Figure \ref{fig:FIG17} illustrates the experimental configuration commonly employed by researchers for investigating series arc faults.

\begin{figure*}[hbt!]
    \centering
    \includegraphics[width=12cm]{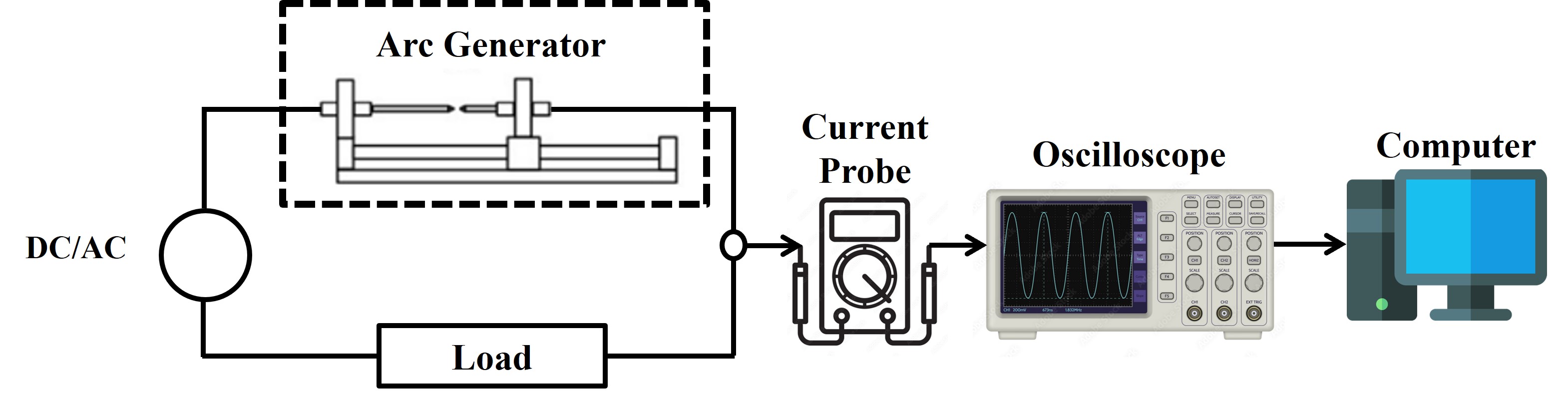}
    \caption{Experimental configuration for a series arc fault. }
    \label{fig:FIG17}
\end{figure*}

 In \cite{dang2021series}, they have two types of scenarios in which test data of a category might be classified. The first scenario occurs when the test data has already undergone training, whereas the second scenario occurs when the test data has not undergone training. The aforementioned scenarios are commonly categorized as enclosed kinds and unenclosed types, respectively. Various artificial intelligence methods, including machine learning and deep learning algorithms, were employed in the study. The results indicated that the performance of different AI systems in arc identification varied significantly. Machine learning algorithms provide strong performance in the low-frequency region; nonetheless, they necessitate the extraction of relevant features. DL algorithms typically require a significant quantity of data in order to achieve optimal performance. The selection of an appropriate technique is of paramount importance in the creation of a dependable and resilient arc fault detection method.

 A novel neural network approach is proposed in \cite{gong2023series}, which incorporates wavelet analysis and feature value decomposition. The data has been divided into five layers using the DB5 wavelet base. The coefficients of the first layer are then extracted in order to generate the Hankel matrix. The initial process of feature extraction and compression has been completed. Upon performing Eigenvalue Decomposition (EVD) on the matrix, the eigenvector is generated, so completing the second phase of feature extraction and data compression. The third stage of feature extraction and data reduction is achieved by computing the mean, root mean square and standard deviation values of the eigenvector. The DNN fault detection model is trained using three input values, with a compressed input scale of 1 × 3. This study introduces a deep neural network (DNN) defect detection model that exhibits low complexity and high timeliness. The complicated nature of the algorithm has been significantly decreased in terms of both operating data and the network's architecture, resulting in a training time of around 33 seconds. The time-efficient model detects faults using current half-cycles. An overall identification rate of 98.7\% was achieved

\subsubsection{Nonlinear Loads}

The method of arc fault detection in the context of nonlinear loads entails the identification of potentially hazardous electrical arcs that may arise within systems operating with nonlinear loads. Nonlinear loads, such as those encountered in contemporary electronic devices, exhibit current consumption patterns that deviate from a straightforward sinusoidal waveform. These loads could generate unpredictable and complex electrical arcs.

The identification of arcs in nonlinear loads necessitates the utilization of sophisticated techniques. One methodology is the examination of the present waveform and the identification of anomalous patterns that signify the presence of an arc. Machine learning and AI algorithms have the capability to undergo training in order to identify these patterns and subsequently activate alarms or implement precautionary measures upon the detection of an arc.
For nonlinear loads, arc fault detection is divided into two types of situations: AC and DC. The researchers' main focus is on low-current circumstances.
The existence of nonlinear loads introduces intricacy to the identification procedure due to the significant variability in their current waveforms. Sophisticated algorithms are required to effectively distinguish between regular load fluctuations and potentially hazardous arc incidents. Examples of nonlinear loads are three-phase motors, inverters, power electronics loads, vacuum cleaners, induction cookers, and similar electrical appliances \cite{han2021recognition,gao2023fault,sung2022tl}. 
Table \ref{tab:non_linear} displays the comparison of approaches employed by various researchers about nonlinear loads.

\begin{itemize}
    \item \textbf{AC Arc Fault Detection}: The adaptive arc identification system has been enhanced and is now accessible, leveraging machine learning techniques. The SVM has strong robustness and possesses the ability to efficiently acquire knowledge and categorize limited sample sizes. Arc fault detection is accomplished by providing spectral characteristics or other relevant parameters as input to the SVM \cite{han2021recognition}. This study presents a recognition approach that utilizes kernel principal component analysis (KPCA) and firefly algorithm optimized support vector machine (FA-SVM). KPCA was employed as a method to effectively distinguish between the harmonics and load noise disturbances present in the voltage and current signals.
The kurtosis and skewness of the 5th and 6th major components were utilized as arc fault characteristics. The FA-SVM was developed with the purpose of detecting and identifying arc faults.
In \cite{9206574}, a series arc fault detection and line selection technique for a multi-load system based on RNN was put forward. The authors developed a method for quick and continuous detection, as well as a method for correcting classification results based on probability which gave them an accuracy of 98.7\%. A novel SAF detection model using a CNN was introduced to efficiently and precisely detect the series arc fault in a three-phase motor with a frequency converter load (TMFCL) circuit. The point-by-point isometric mapping technique was introduced as a means to generate the input matrix. The model's lightweight architecture was achieved through the utilization of constraint building blocks and depth-wise separable convolutions. The level of difficulty and expected performance of convolution operators was conducted using a roofline model. Based on an analysis of the operators' runtime, the researchers in \cite{wang2023line} identified and designated the most efficient lightweight SAF identification model as SAFNet. In \cite{9431207} A novel approach was proposed for detecting series arc faults and selecting phases, utilizing single-phase current. The integration of the Fractional Fourier Transform (FRFT) and the two-level block SVD is highly effective in extracting arc fault detection and achieving phase selection. 
A way for finding arc fault that combines the channel attention mechanism and the residual network model is outlined in \cite{gao2023fault}. This research presents a novel module called one-dimensional depth separable convolution (1D-DS) that aims to decrease the number of parameters in a network model and lower the inference time for individual prediction samples. The accuracy of their model was found to be 98.07\%.

\item \textbf{DC Arc Fault Detection}: The authors in \cite{sung2022tl} put forward a two-stage training strategy for a transfer learning-based low-energy arc-fault detection network (TL-LEDarcNet) to identify series DC arc-faults. A lightweight one-dimensional convolutional neural network and a one-layer LSTM network were created to detect low-energy arc faults using only sensed current information and got an accuracy of 95.8\%. 
\end{itemize}

\begin{table}
\caption{Comparison of methodologies used for nonlinear loads.}
\centering
\renewcommand{\arraystretch}{1.5}
\begin{tabular}{c c l c c}
\toprule
  \multicolumn{1}{>{\centering}m{3em}}{Reference} &
  \multicolumn{1}{>{\centering}m{3em}}{Load} &
  \multicolumn{1}{l}{Methodology} &
  \multicolumn{1}{>{\centering}m{3em}}{Accuracy} &
 \\
  \midrule
  \cite{han2021recognition}   & 3   & KPCA and FA-SVM  & 97.17\%  \\ 
\cite{9206574}                & 1   & RNN              & 98.70\%  \\ 
\cite{wang2023line}           & 1   & CNN              & 99.44\%  \\ 
\cite{9431207}                & 8   & SVM and PSO      & 97.53\%  \\ 
\cite{gao2023fault}           & 6   & CNN              & 98.07\%  \\ 
\cite{sung2022tl}             & 1   & LSTM             & 95.80\%  \\ 

\bottomrule
\end{tabular}
\label{tab:non_linear}
\end{table}

\subsubsection{Linear Loads}
Linear loads refer to electrical devices that exhibit a direct proportionality between the current flowing through them and the voltage applied across them. Examples of linear loads are incandescent lights, resistive heaters, and similar electrical appliances. The primary difficulty associated with arc fault identification in the presence of linear loads is the differentiation between typical electrical variations and hazardous arc faults that have the potential to cause fires or pose other safety risks. The identification of characteristic waveforms or patterns in electrical signals is of utmost importance in the early detection of arc faults.Table \ref{tab:linear} displays the comparison of approaches employed by various researchers about linear loads or no load.

\begin{itemize}
    \item \textbf{DC Arc Fault Detection}: In \cite{miao2020dc}, the system utilizes arc time-frequency signatures obtained by a modified empirical mode decomposition (EMD) technique and employs the SVM method for making decisions. The current data in the time domain for load arc faults is subjected to multilayer discrete wavelet analysis. A coefficient matrix is created from five-layer discrete wavelet detail values and translated into an RGB phase space image using a colormap index. They introduced wavelet compression reconstruction-based data improvement.
    \item \textbf{AC Arc Fault Detection}: A novel method for diagnosing arc defects is presented that incorporates SVM, improved multi-scale fuzzy entropy (IMFE), and variational mode decomposition (VMD) techniques  \cite{wang2021arc}. Using VMD, the current signal is decomposed into intrinsic mode functions (IMFs) in the time-frequency domain. The associated IMFs with the arc defect are then utilized to calculate IMFE. Finally, SVM detects IMFEs utilized for arc faults. The article \cite{lala2020detection} focuses on the examination of two primary events: the arc results from a tree's inclination on a medium-voltage distribution line, as well as the arc that arises between sphere gaps. The examination of real-time arc signals is conducted using a combined approach of EMD and ANN. The utilization of EMD and ANN in the analysis of arc voltage data has proven to be effective in accurately identifying and categorizing arcing occurrences based on their main harmonic characteristics. They got their highest accuracy of 100\% and lowest accuracy of 98.2\%.
\end{itemize}

\begin{table}
\caption{Comparison of methodologies used for linear loads and no load.}
\centering
\renewcommand{\arraystretch}{1.5}
\begin{tabular}{c c l c c}
\toprule
  \multicolumn{1}{>{\centering}m{4em}}{References} &
  \multicolumn{1}{>{\centering}m{6em}}{Load Type} &
  \multicolumn{1}{l}{Methodology} &
  \multicolumn{1}{>{\centering}m{3em}}{Accuracy} &\\
  \midrule
  \cite{miao2020dc}  &  Resistive          & EMD and SVM  &  96.00\%  \\ 
 \cite{wang2021arc}                & Resistive         & 
VMD, IMFE, and SVM & 99.00\%  \\ 
\cite{lala2020detection}      & No Load        & EMD and ANN    & 98.20\%  \\ 
\bottomrule
\end{tabular}
\label{tab:linear}
\end{table}

It is worth noting that there are specific research that demonstrates a higher level of accuracy in detecting DC arc faults. However, it is evident that there is a lack of similar discussions on AC arc faults. In the same way, while many researchers have engaged in discussions on the accuracy pertaining to low-voltage situations, a corresponding examination within the framework of high-voltage settings is conspicuously lacking in the existing body of research.
{Integrating AI in real-world electrical distribution systems presents significant potential to improve efficiency, precision, and service dependability. The increased complexity of designing and operating power systems makes AI immensely significant for the power industry\cite{HEYMANN2024100322}. Despite initial expenses, investing in AI infrastructure and deployment establishes a platform for enduring advantages as these technologies expand across various locations. While legacy infrastructure may require modifications, these enhancements enable smooth AI integration, encouraging interoperability and advanced data insights. Scaling AI models across distribution networks poses the valuable issue of improving model performance for comprehensive real-world applications, and adaptive models can continually learn and evolve with system changes. Regulatory compliance, particularly about data privacy and transparency, creates trust and protects consumer interests, boosting the trustworthiness and acceptability of AI-driven solutions. Additionally, ethical principles around justice and accountability integrate AI deployments with current governance expectations, increasing these technologies' integrity and societal effect\cite{MALIK2024102886}. Together, these efforts lead to more resilient, efficient, and accountable electrical distribution networks, placing AI as a transformational force in electrical distribution systems.}

\begin{table}[ht]
\caption{Comparison of Arc Fault Detection Methods}
\centering
\renewcommand{\arraystretch}{1.2} 
\setlength{\tabcolsep}{5pt} 
\resizebox{\linewidth}{!}{ 
\begin{tabular}{p{3.5cm} p{2cm} p{3.5cm} p{2.5cm} p{1.5cm}}
\toprule
\textbf{Methodology} & \textbf{Type} & \textbf{Data source and Type of system} & \textbf{Type of Data} & \textbf{Accuracy} \\
\midrule
Current amplitude spectrum and the sparse representation algorithm \cite{qu2018arc} & Traditional & Experimental data and LVAC system & Current waveforms & 100\% \\ 
Noninvasive sensor-based method \cite{7012066} & Traditional & Experimental data and LVAC system & EMR signals and current signals & 100\% \\ 
Chirp Zeta Transform (CZT) \cite{7814140} & Traditional & Experimental data and LVAC system & Current signals & - \\ 
CNN \cite{wang2021arcnet} & AI-Based & Experimental data and LVAC system & Current signals & 99.47\% \\ 
Random Forest (RF) - DNN \cite{jiang2021series} & AI-Based & Experimental data and LVAC system & Current waveforms & 97.50\% \\ 
Composite bandpass filter (CBF), principal component analysis (PCA), temporal convolution network (TCN) \cite{10054597} & AI-Based & Experimental data and LVDC system & Current signals & 99.88\% \\ 
\bottomrule
\end{tabular}
}
\label{tab:comparision}
\end{table}

Table \ref{tab:comparision} demonstrates the comparison between traditional and AI-based arc fault detection systems. Traditional techniques attained an exceptional 100\% accuracy, emphasizing the trustworthiness in experimental LVAC systems employing current waveforms and electromagnetic radiation (EMR) data. However, these technologies frequently encounter difficulties in adaptation and scalability in complex and dynamic systems. On the other hand, AI-driven techniques have shown remarkable potential, with accuracy rates approaching perfection: 99. 47\% for CNN-based detection, 97.50\% for RF integrated with deep neural networks (DNN), and an exceptional 99. 88\% for CBF using PCA and TCN. These AI-driven methodologies excel in processing extensive data sets, providing enhanced adaptability, accuracy, and the capacity to generalize across many systems, including LVAC and LVDC. Their ability to independently learn intricate patterns renders them as highly scalable, resilient, and effective solutions for real-time fault detection, representing a notable improvement over conventional methods in contemporary power systems.
We have done a methodological investigation of arc faults. The article presents a detailed assessment of numerous methodologies for identifying arc defects, including traditional methods and AI-driven strategies. It explains the principles behind these technologies, their advantages, limitations, and the efficiency of alternative detection procedures. The study seeks to offer a complete review of the improvements in arc fault detection strategies, highlighting the value of both conventional and AI-based approaches in boosting detection accuracy and efficiency.
\subsection{Limitations of Artificial Intelligence and Machine Learning Approaches}

The utilization of AI and ML in the domain of arc fault detection is subject to several constraints, which include difficulties pertaining to the quality of data, intricacy of models, ability to generalize, interpretability, and the requirement for continuous model maintenance. 
The restricted interpretability of outcomes is dependant upon the quality and amount of the available data. In addition, the system encounters several hurdles in effectively resolving complex or uncommon problems, is vulnerable to the presence of irrelevant or extreme data points, and experiences difficulties in determining appropriate algorithms and parameters\cite{9018242}. The transmission of knowledge between disparate areas poses a significant challenge. The reliance on substantial quantities of precisely annotated data is a crucial determinant \cite{CHEN2022108018}. The issue of sensitivity to hyperparameter selection holds great importance. Within the domain of machine learning, a significant issue arises about the possibility of algorithms becoming overfitted to the training data \cite{wang2018series}. The recognition of these limits is of utmost relevance for stakeholders operating in safety-critical industries, as it provides guidance for the development and deployment of robust and effective arc fault detection systems.

{We acknowledge that the demands for processing resources and data constraints are significant obstacles in implementing AI for this application. To alleviate the computational burden, we advocate for implementing edge computing \cite{bdcc8080094}, which processes data nearer to its origin, hence reducing latency and improving response times in detecting arc faults. In addition, model compression methodologies—such as pruning, quantization, and knowledge distillation—facilitate the reduction of size and complexity of AI models, enhancing their applicability for deployment on resource-limited devices. These technologies improve efficiency while mitigating data constraints, enabling the deployment of robust AI models without the need for excessive data or computational resources\cite{sinha2024challenges}. Combining these methodologies will strengthen our study, demonstrating realistic pathways to overcome present AI limits in arc fault identification.}

\section{Conclusion}
\label{section:Conclusion}
The aim of this systematic study is to offer a detailed examination of the most recent advancements in approaches for detecting arc faults. This review comprises many approaches, including time-domain examination, frequency domain examination, and time-frequency domain examination, alongside the analysis of physical events and the utilization of artificial intelligence. The present review centers on the incorporation of AI in the context of arc fault detection. The integration of AI methods has demonstrated various benefits, such as the capacity to analyze substantial volumes of data in real time with enhanced precision, quicker reaction times in the detection of arc faults, and the potential to learn
 for identifying particular fault types.

Despite the numerous advantages associated with artificial intelligence, it is essential to acknowledge and address the limitations that exist within current arc fault detection approaches. The implementation of deep learning methods on embedded devices is often challenging due to the intricate network architecture, numerous parameters, and extensive computational requirements. To achieve optimal accuracy, it is important to construct an extensive database. To address the issue of limited data availability, it is imperative to create effective techniques for data gathering and preparation. Fault detection techniques that rely on data are vulnerable to anomalies caused by factors for instance measurement problems, disruptions caused by machines, and the start-up procedure for loads. These anomalies could lead to the generation of unfavourable data, which hinders the accurate detection of abnormalities.

AI methods have demonstrated superior accuracy compared to other approaches. Nevertheless, it is important to acknowledge that these techniques necessitate a substantial volume of data for the training process and rely on significant computational resources. In contrast, classic methods such as the over-current and wavelet-based approaches offer advantages in terms of simplicity and speed, respectively. Hybrid methods have been developed as an initiative to integrate the benefits of different approaches. These procedures are characterized by their increased complexity, yet they yield exceptional outcomes.

As the volume and quality of data rises, the effectiveness of data-driven processes is enhanced. In cases where a system possesses a comprehensive and dependable model, model-based approaches would be more advantageous alternatives. This study facilitates the user in identifying the most efficient approach for fault detection and exploring different techniques to enhance the system's capacity.

 Further investigation should be conducted into the utilization of synthetic data generation and transfer learning techniques as a means to enhance the training dataset. Furthermore, the combination of various artificial intelligence methodologies and the incorporation of expert knowledge have the potential to enhance the precision of fault detection and mitigate the occurrence of false alarms. It is worth studying algorithms that could effectively and efficiently operate on hardware with limited processing capability. In conclusion, it could be inferred that artificial intelligence holds significant potential in changing the domain of arc fault detection, providing a beneficial and precise methodology to address the crucial issue of safeguarding electrical safety.

\bibliographystyle{IEEEtran}
\bibliography{main}

\end{document}